\documentclass[manuscript,nonacm,screen]{acmart}

\usepackage{tikz}
\usetikzlibrary{cd,positioning,calc,arrows}
\usepackage{graphicx,latexsym,alltt}
\usepackage{url}
\newcommand{\proofrule}[2]{\displaystyle\frac{#1}{#2}}
\newcommand{\Forall}[1]{\forall #1\,{\cdot}\,}
 
\newcommand{\Space}{\hspace{0.3cm}}

\newcommand{\Sig}{\mathtt{Sig}}
\newcommand{\Mod}{\mathtt{Mod}}
\newcommand{\Sen}{\mathtt{Sen}}

\newcommand{\LABEL}{\mathtt{LABEL}}
\newcommand{\Label}{\mathtt{Label}}

\newcommand{\QUEUE}{\mathtt{QUEUE}}

\newcommand{\OMEGA}{\mathtt{OMEGA}}

\newcommand{\BOOL}{\mathtt{BOOL}}
\newcommand{\Bool}{\mathtt{Bool}}

\newcommand{\TRIV}{\mathtt{TRIV}}
\newcommand{\Elt}{\mathtt{Elt}}
\newcommand{\NAT}{\mathtt{NAT}}
\newcommand{\PNAT}{\mathtt{PNAT}}
\newcommand{\Nat}{\mathtt{Nat}}

\newcommand{\EX}{\mathtt{EX}}
\newcommand{\ID}{\mathtt{ID}}
\newcommand{\US}{\mathtt{US}}

\newcommand{\A}{\mathcal{A}}
\newcommand{\B}{\mathcal{B}}
\newcommand{\M}{{\mathcal{M}}}

\newcommand{\N}{{\mathbb{N}}}

\newcommand{\red}{\upharpoonright}

\newcommand{\SP}{\mathtt{SP}}
\newcommand{\IH}{\mathtt{IH}}

\newcommand{\codesize}{\small}

\newcommand{\mi}[1]{\mathit{#1}}

\title{Proof Scores: A Survey (full version)}

\author{Adri{\'{a}}n Riesco}
\orcid{0000-0002-9716-4612}
\affiliation{\institution{Universidad Complutense de Madrid}
  \city{Madrid}
  \country{Spain}}
\email{ariesco@fdi.ucm.es}
\author{Kazuhiro Ogata}
\orcid{0000-0002-4441-3259}
\affiliation{\institution{Japan Advanced Institute of Science and Technology}
  \city{Nomi}
  \country{Japan}}
\email{ogata@jaist.ac.jp}
\author{Masaki Nakamura}
\affiliation{\institution{Toyama Prefectural University}
  \city{Imizu}
  \country{Japan}}
\email{masaki-n@pu-toyama.ac.jp}
\author{Daniel G{\u{a}}in{\u{a}}}
\orcid{0000-0002-0978-2200}
\affiliation{\institution{Kyushu University}
  \city{Fukuoka}
  \country{Japan}}
\email{daniel@imi.kyushu-u.ac.jp}
\author{Duong Dinh Tran}
\orcid{0000-0001-7092-2084}
\affiliation{\institution{Japan Advanced Institute of Science and Technology}
  \city{Nomi}
  \country{Japan}}
\email{duongtd@jaist.ac.jp}
\author{Kokichi Futatsugi}
\affiliation{\institution{Japan Advanced Institute of Science and Technology}
  \city{nomi}
  \country{Japan}}
\email{futatsugi@jaist.ac.jp}

\begin{abstract}
Proof scores can be regarded as outlines of the formal verification of system properties.
They have been
historically used by the OBJ family of specification languages.
The main advantage of proof scores is that they follow the same syntax
as the specification language they are used in, so specifiers can
easily adopt them and use as many features as the particular language 
provides. In this way, proof scores have been successfully used to prove 
properties of a large number of systems and protocols.
However, proof scores also present a number of disadvantages that prevented
a large audience from adopting them as proving mechanism.

In this paper we present the theoretical foundations of proof scores; 
the different systems where they have been adopted and
their latest developments;
the classes of systems successfully verified using proof scores, including the
main techniques used for it;
the main reasons why they have not been widely adopted; and finally we discuss
some directions of future work that might solve the problems discussed previously.

\end{abstract}

\ccsdesc[500]{Theory of computation~Proof theory}
\ccsdesc[300]{Security and privacy~Logic and verification}

\ccsdesc[500]{Software and its engineering~Formal methods}
\ccsdesc[500]{Software and its engineering~Software verification} 

\definecolor{duongColor}{rgb}{0.24, 0.71, 0.54}

\keywords{CafeOBJ, Theorem proving, System specification, rewriting}

\begin{document}

\maketitle

\section{Introduction}\label{sec:intro}

This paper is a full version of the published paper~\cite{cafe-survey}, including supplemental materials.

Proof scores~\cite{proofScores} were originally proposed in the 90's as a promising technique for proving properties of systems using term rewriting. 
The development of algebraic specification languages executable by rewriting, and in particular
of the OBJ family of programming languages, allowed specifiers to put this methodology into practice, revealing proof scores as a powerful   alternative to standard theorem proving approaches.

The key feature of proof scores is its complete integration into the specification process:  
proof scores use the same syntax and evaluation mechanism as the specification language, hence easing the verification process.
This smooth integration of the specification methodology into the verification process has allowed practitioners to analyze several systems and protocols over the last decades.
However, there are some weak points since the efforts were directed to the development of new features for the specification languages executable by rewriting assuming that proof scores would benefit from these new features (as it happens), but
leaving the verification burden such as checking that it follows ``sound rules,'' to the users.

Taking these ideas into account, and despite their good properties, 
the use of proof scores has been mostly limited to academic environments,
without reaching a wide audience, in particular software development
companies.
Although this lack of industrial applications is not limited to proof scores,
it is worth analyzing the reasons why this happens and how it could be solved.

In this paper we discuss the past, present, and future of proof scores.
In particular, for the past we describe their history,
as well as most of the systems and protocols that have been specified and
verified thus far,
which can be used as a bibliography by topic.
For the present, 
we give an updated reference guide to work with them, 
discuss their theoretical basis,
and explain in detail a protocol that
illustrates some directions to consider in the future.
In turn, we analyze past and present to understand the problems
and challenges when using proof scores and suggesting the most
promising lines of future work.

The rest of the paper is organized as follows: Section~\ref{hn_sect} briefly
introduces the history of proof scores. 
Section~\ref{sec:theory} presents the theoretical ideas of proof scores,
while Section~\ref{sec:spec} introduces structured specifications and their application to proof scores.
Section~\ref{sec:tools} lists the tools supporting proof scores, discussing their different features.
Section~\ref{sec:success} presents proof scores in practice, illustrating how
to use proof scores using a running example. Moreover,
it also describes complementary verification techniques that complement proof
scores and ease its use.
Section~\ref{sec:success_cases} outlines most of the types of protocols successfully 
analyzed using proof scores, giving particular examples of each category and
discussing in detail one protocols that illustrates how state-of-the-art protocols 
are analyzed.
Section~\ref{sec:open} discusses the open issues and challenges proof scores
face in the future and possible ways to deal with them, while 
Section~\ref{sec:rel}
presents the related work and its relation with proof scores.
Finally, Section~\ref{sec:conc} concludes and details lines of ongoing and future work.
The examples shown in the survey are available at
\url{https://github.com/ariesco/proof-scores-survey}

\section{History in a Nutshell}\label{hn_sect}

The terminology ``Proof Score'' was coined by Joseph
A. Goguen~\cite{Goguen90ALP}, one of the first researchers to apply
Category Theory (CT) to Computer Science (CS) and one of the pioneers
in the field of Algebraic Specification. His PhD
work~\cite{Goguen68PhD} is on categories of Fuzzy sets called Goguen
Categories~\cite{Winter2007TL}.  As another application of CT to CS,
Goguen, together with Burstall, invented the Theory of
Institutions~\cite{GoguenB92JACM}, which was born as
the theoretical foundation of parameterized
programming~\cite{Goguen84TSE} and then came to be used as a
framework to formalize logics. Goguen and Burstall designed
Clear~\cite{BurstallG79ASS}, a specification language, based on the
Theory of Institutions. Although Clear was not implemented, Goguen,
together with Futatsugi, Jouannaud, and Meseguer, designed and
implemented OBJ2~\cite{DBLP:conf/popl/FutatsugiGJM85}, an algebraic
specification language, which was followed by
OBJ3~\cite{Goguen99SEOBJ}. OBJ3 is one of the earliest computer
languages that made it possible to do parametrized programming. One
key concept of Clear and OBJ3 to support parametrized programming is
pushout-based parametrized modules, which take modules as their
parameters. Clear and OBJ3 have influenced module systems in some
programming languages, such as Ada, Standard ML, and
C++~\cite{Goguen97UPub}. Standard ML modules~\cite{MacQueen84LISP}
were based on prototype designs of modules for
Hope~\cite{BurstallMS80LISP}, a functional programming language,
which were influenced by Clear.  Among the other earliest computer
languages that supported parameterized programming are ACT
ONE~\cite{EhrigFH83ADT} and
HISP~\cite{DBLP:conf/ifip/FutatsugiO80}. 
OBJ3 was succeeded by
CafeOBJ\footnote{\url{https://cafeobj.org/}}~\cite{DiaconescuF98WS}
and
Maude\footnote{\url{http://maude.cs.illinois.edu/w/index.php/The_Maude_System}}~\cite{maude-book}. Thus,
CafeOBJ and Maude are sibling languages. OBJ3 was implemented in
Common Lisp and so was CafeOBJ, while Maude was implemented in
C++. 
In contrast to OBJ3 and Maude, which have one implementation, CafeOBJ has
two implementations. 
The second implementation~\cite{cafeInMaudeFAC}
of CafeOBJ, called CafeInMaude, was conducted in Maude, which can be used as logical framework
as well. 

Goguen~\cite{Goguen90ALP} describes proof scores as follows:

\begin{quote}
The basic idea is to transform theorem proving problems into {\bf
  proof scores}, which consist of declarations and rewritings such
that if everything evaluates as desired, then the problem is
solved. This approach is neither fully automatic nor fully manual, but
rather calls for machines to do the most routine tasks, such as
substitution, simplification and reduction, and for humans to do the
most interesting tasks, such as deciding proof strategies. Moreover,
partially successful proofs often return information that suggests
what to try next, \ldots
\end{quote}

\noindent
Goguen and his colleagues initially used proof scores to tackle theorem
proving problems about data structures, such as natural numbers. There
was criticism about algebraic specification
techniques, which are based on elegant theories but do not have any
practical applications. Eric G. Wagner mentions such criticism in his
article~\cite{Wagner02NJC} about algebraic specifications. Wagner was
a member of the ADJ group~\cite{Goguen93CTTCS,Wagner02NJC} and so was
Goguen. The criticism has been addressed by introducing hidden algebra
and observational specifications~\cite{hiddenAlgebra,DiaconescuF00JUCS}
and the work of Futatsugi, Ogata, and others, who have 
demonstrated that proof scores could
be used to tackle theorem proving problems about a wide range of
systems, among which are authentication protocols and electronic
commerce protocols (see Section~\ref{sec:success_cases} for details).

There was criticism specific to proof scores as well. This is because
humans are supposed to do the most interesting tasks, such as deciding
proof strategies. Thus, proof scores are subject to human errors. This
criticism has been addressed by developing proof assistants, such as the
Inductive Theorem Prover (ITP)~\cite{DBLP:journals/jucs/ClavelPR06} and the
Constructor-based Inductive Theorem Prover (or
CITP)~\cite{DBLP:conf/calco/GainaZCA13,gainaTR18}. Following the development of CITP, a
proof assistant was developed inside CafeOBJ and the CafeInMade Proof
Assistant (CiMPA)~\cite{tosem18} was developed for
CafeInMaude. ITP, CITP, the CafeOBJ proof assistant, and CiMPA can prevent
human errors but may dilute the merits of proof scores, although those
proof assistants generate proof scores inside. To address the issue,
Riesco and Ogata developed an automatic proof script generator for
CiMPA from manually written proof scripts in CafeOBJ, called the CafeInMaude Proof Generator (CiMPG)~\cite{tosem18}.
 
\section{Theoretical foundations}\label{sec:theory}

In this section we discuss the fundamental theoretical concepts required to understand how the proof score approach works. 
The discussion is limited to basic order-sorted specifications even though both Maude and CafeOBJ (see Section~\ref{sec:tools}) 
are based on more advanced logical semantics which support, for example, rewriting or behavioural specifications.

\paragraph{\bf Signatures}
We introduce \emph{order-sorted signatures}, which describe the relations between \emph{sorts} and how terms are built.
The main ideas underlying these signatures are \emph{inheritance} and \emph{polymorphism}.

\begin{definition}[Order-sorted signatures~\cite{orderSorted}]
A \emph{many-sorted signature} is a pair $(S, F)$ where $S$ is a set of sorts and 
$F=\{\sigma:w\to s\mid w\in S^* \text{ and } s\in S\}$ is a set of function symbols.
For each function symbol $\sigma:w\to s$ in $F$, we say that $\sigma$ is its name, $w$ is its arity, and $s$ is its co-arity.
Notice that $F$ can be regarded as an $(S^* \times S)$-sorted set $\{F_{w,s}\}_{(w,s)\in S^*\times S}$, 
where $F_{w,s}=\{\sigma\mid \sigma:w\to s \in F\}$. 
An order-sorted signature is a triple $(S, \leq, F)$ such that $(S, F)$ is a  many-sorted signature, and $(S, \leq)$ is a partially ordered set.
\end{definition}

A function symbol $\sigma \in F_{\lambda, s}$ with the empty arity $\lambda$ is called a \emph{constant}.
The set $S/_{\equiv_\leq}$ of connected components of $(S, \le)$ is the quotient of $S$ under the equivalence relation $\equiv_\leq$ generated by $\leq$.
The equivalence relation $\equiv_\leq$ is canonically extended to strings over $S$.
The set of ground terms $\mathcal{T}_\Sigma$ over a signature $\Sigma=(S,\leq,F)$ is defined as follows:
$$\begin{array}{l l l}
1)~\displaystyle\frac{}{c\in \mathcal{T}_{\Sigma,s}} \text{ if } c:~\to s\in F & 
2)~\displaystyle\frac{t_1\in \mathcal{T}_{\Sigma,s_1} ~\dots~  t_n\in \mathcal{T}_{\Sigma,s_n} }{\sigma(t_1,\dots,t_n)\in \mathcal{T}_{\Sigma,s}} \text{ if } \sigma:s_1\dots s_n \to s\in F &
3)~\displaystyle\frac{t\in \mathcal{T}_{\Sigma,s_0}}{t\in \mathcal{T}_{\Sigma,s}} \text{ if } s_0\leq s
\end{array}
$$
We denote by $\mathcal{T}_\Sigma(X)$ the set of terms with variables from $X$. 

\begin{definition}[Regularity]
A signature $\Sigma=(S,\leq,F)$ is \emph{regular} iff for any function symbol $\sigma:w\to s\in F$ and any string of sorts $w_0 \in S^*$ such that $w_0 \leq w$, the set $\{(w',s') \in S^*\times S \mid w_0\leq w' \text{ and }\sigma:w'\to s'\in F\}$ has a least element.
\end{definition}
By regularity each term has a least sort.
In practice, we work only with regular signatures.

\begin{example}[Natural numbers] \label{ex:sig-nat}
Let $\Sigma_\NAT$ be the following signature of natural numbers:
\begin{enumerate}
\item The set of sorts is $\{\mathit{Nat}, \mathit{NzNat}\}$ such that $\mathit{NzNat} \leq \mathit{Nat}$.
\item The set of function symbols is $F=\{0 : ~ \to \mathit{Nat}, s\_: \mathit{Nat} \to \mathit{NzNat}, \texttt{\_+\_}: \mathit{Nat}~\mathit{Nat}\to \mathit{Nat} \}$.
\end{enumerate}
\end{example}

Clearly, the signature of Example~\ref{ex:sig-nat} is regular.

\paragraph{\bf Models}
The models consist of order-sorted algebras which give the denotational semantics of specifications.

\begin{definition}[Order-sorted algebras~\cite{orderSorted}]
Let $\Sigma=(S, \leq, F)$ be an order-sorted signature.

An order-sorted algebra $\A$ over $\Sigma$ consists of an $(S,F)$-algebra, that is,  
\begin{itemize}
\item an $S$-sorted set $\{\A_s\}_{s\in S}$, and
\item a function $\A_\sigma:\A_w\to\A_{s}$ for each function symbol $\sigma:w\to s\in F$, 
where $\A_w=\A_{s_1}\times\dots\times \A_{s_n}$ whenever $w=s_1\dots s_n$ and $\A_w$ is a singleton whenever $w$ is the empty string,
\end{itemize}

satisfying the following properties:
\begin{itemize}
\item $\A_s \subseteq \A_{s'}$ whenever $s\leq s'$, and

\item for all $\sigma: w_1\to s_1 \in F$ and $\sigma:w_2\to s_2\in F$ such that $w_1\equiv_\leq w_2$,
the functions $\A_\sigma : \A_{w_1} \to \A_{s_1}$ and $\A_\sigma : \A_{w_2} \to \A_{s_2}$ return the same value for the same arguments (in $\A_{w_1}\cap \A_{w_2}$).
\end{itemize}
An order-sorted homomorphism $h:\A\to\B$ is a many-sorted homomorphism such that $h_s(a)=h_{s'}(a)$ for all sorts $s,s'\in S$ with $s\equiv_\leq s'$ and all elements $a\in \A_s\cap\A_{s'}$.
We denote by $\Mod(\Sigma)$ the category whose objects are order-sorted $\Sigma$-algebras and arrows are order-sorted $\Sigma$-hommorphisms.
\end{definition}

The order-sorted algebra of natural numbers $\mathcal{N}$ over $\NAT$ interprets 
$\mathit{Nat}$ as the set of all natural numbers, 
$\mathit{NzNat}$ as the set of all positive natural numbers,
and $0$ as zero, 
$s\_$ as the successor function, and
$\texttt{\_+\_}$ as the addition.

Clearly, $\mathcal{T}_\Sigma$ can be organized as an order-sorted algebra, interpreting each function symbol $\sigma:s_1\dots s_n\to s\in F$ as a function $\mathcal{T}_\sigma:\mathcal{T}_{s_1}\times\mathcal{T}_{s_n}\to \mathcal{T}_s$ defined by $\mathcal{T}_\sigma(t_1,\dots,t_n)=\sigma(t_1,\dots,t_n)$ for all $i\in\{1,\dots,n\}$ and $t_i\in \mathcal{T}_{s_i}$.
Similarly, for any set of variables $X$ for $\Sigma$, the set of terms with variables $\mathcal{T}_\Sigma(X)$ can be organized as a $\Sigma$-algebra.

\paragraph{\bf Sentences}
The sentences consist of universally quantified conditional equations which make the specifications executable by rewriting.

\begin{definition}[Conditional equations]
The sentences over a signature $\Sigma=(S,\leq,F)$ are (conditional) equations of the form $\Forall{X}t_0=t_0'\mbox{ if } t_1 = t_1'\wedge \ldots\wedge t_n=t_n'$, 
where $n\in\N$,  
$X$ is a finite set of variables for $\Sigma$, and 
$t_i$ and $t_i'$ are terms with variables from $X$ such that the sorts of $t_i$ and $t_i'$ are in the same connected component.
We denote by $\Sen(\Sigma)$ the set of $\Sigma$-sentences.
\end{definition}
If $n=0$ then the sentence is an equation $\forall{t_0=t_0'}$. 

\begin{example}[Addition]\label{ex:eq-nat}
Some examples of sentences over the signature $\Sigma_\NAT$ are the equations which inductively define the addition of natural numbers,
$E_\NAT\coloneqq \{\Forall {x:\mathit{Nat}} 0 + x = x,~\Forall{x,y:\mathit{Nat}}s~y + x = s(x + y)\}$.
\end{example}

\paragraph{\bf Satisfaction relation}
The satisfaction of a sentence by an order-sorted algebra is the usual Tarskian satisfaction based on the interpretation of terms.
More concretely, given a signature $\Sigma$, an order-sorted $\Sigma$-algebra $\A$ satisfies a $\Sigma$-sentence $\Forall{X}t_0=t_0'\mbox{ if } t_1 = t_1'\wedge \ldots\wedge t_n=t_n'$, 
in symbols,
$\A\models_\Sigma \Forall{X}t_0=t_0'\mbox{ if } t_1 = t_1'\wedge \ldots\wedge t_n=t_n'$, 
iff for all evaluations $f:X\to \A$ we have $f^\#(t_0)=f^\#(t'_0)$ whenever $f^\#(t_i)=f^\#(t'_i)$ for all $i\in\{1,\dots,n\}$, 
where $f^\#:\mathcal{T}_\Sigma(X)\to \A$ is the unique homomorphism extending the valuation $f:X\to\A$.
The satisfaction relation is straightforwardly generalized to sets of sentences:
$\A\models_\Sigma E$ iff  $\A\models_\Sigma e$ for all $e\in E$.
We say that a set of sentences $E$ \emph{satisfies} a set of sentences $E'$, in symbols, $E\models_\Sigma E'$, iff $\A\models_\Sigma E$ implies $\A\models_\Sigma E'$, for all $\Sigma$-algebras $\A$.
Notice that the algebra of natural numbers $\mathcal{N}$ with addition defined above satisfies both equations from Example~\ref{ex:eq-nat},
since $0 + n = n$ for all $n\in\N$ and $\mathcal{N}_s~m + n =\mathcal{N}_s(m+n)$ for all natural numbers $m,n\in\N$.

\paragraph{\bf Signature morphisms}
A very important topic in algebraic specifications, in particular, and in computer science, in general, is modularization.
One crucial concept for modularization is the notion of signature morphism.

\begin{definition}\label{def:sig-morph}
A signature morphism $\chi\colon (S,\leq,F)\to (S',\leq',F')$ 
is a many-sorted algebraic signature morphisms $\varphi\colon (S,F)\to(S',F')$ 
such that the function $\chi\colon (S,\leq)\to(S',\leq')$ is monotonic and 
$\chi$ preserves the subsort polymorphism, that is,
$\chi$ maps each function symbols with the same name $\sigma:w_1\to s_1$ and $\sigma:w_2\to s_2$ and such that $w_1\equiv_\leq w_2$ to some function symbols $\sigma’:w_1’ \to s_1'$ and $\sigma’: w_2'\to s_2'$ that have the same name. 
\end{definition}

In Definition~\ref{def:sig-morph}, 
since $\chi\colon (S,\leq)\to(S',\leq')$ is monotonic, 
$w_1\equiv_\leq w_2$ implies $\chi(w_1)\equiv_{\leq'}\chi(w_2)$. 
In practice, most of signature morphisms used are inclusions, which obviously preserves the subsort polymorphism.

\begin{example}\label{ex:triv-nat}
Let $\Sigma_\TRIV$ be the signature which consists of one sort $\mathit{Elt}$, and
$\chi:\Sigma_\TRIV \to \Sigma_\NAT$ be the signature morphism which renames $\mathit{Elt}$ to $\mathit{Nat}$. 
\end{example}

\begin{example} \label{ex:nat-list}
Let $\Sigma_\mathtt{NATL}$ be the signature of lists of natural numbers which extends the signature of natural numbers $\Sigma_\NAT$ with
the sorts $\{\mathit{NeList},\mathit{List}\}$ such that $\mathit{Nat} < \mathit{NeList}< \mathit{List}$, and 
the following function symbols.
\begin{center}
\begin{tabular}{l l l}
\begin{minipage}{0.3\textwidth}
\begin{itemize}
\item $\mathit{empty} : ~ \to \mathit{List}$
\item $\texttt{\_;\_}: \mathit{List}~\mathit{List}\to \mathit{List}$
\end{itemize}
\end{minipage}
&
\begin{minipage}{0.3\textwidth}
\begin{itemize}
\item $\texttt{\_;\_} : \mathit{NeList}~\mathit{List}\to \mathit{NeList}$
\item $\texttt{\_;\_} : \mathit{List}~\mathit{NeList}\to \mathit{NeList}$
\end{itemize}
\end{minipage}
&
\begin{minipage}{0.3\textwidth}
\begin{itemize}
\item $\mathit{tail} : \mathit{NeList} \to \mathit{List}$
\item $\mathit{head}: \mathit{NeList} \to \mathit{Nat}$
\end{itemize}
\end{minipage}
\end{tabular}
\end{center}
The inclusion $\iota:\Sigma_\NAT\hookrightarrow \Sigma_{\mathtt{NATL}}$ is another example of signature morphism.
\end{example}
Any signature morphism $\chi:\Sigma\to \Sigma'$ can be extended to a function $\chi:\Sen(\Sigma)\to\Sen(\Sigma')$ which translates the sentences over $\Sigma$ along $\chi$ in a symbolwise manner.
On the other hand, for any signature morphism $\chi:\Sigma\to \Sigma'$, there exists a reduct functor $\red_\chi\_:\Mod(\Sigma')\to\Mod(\Sigma)$ defined by:
\begin{itemize}
\item For all $\A'\in|\Mod(\Sigma')|$, $(\A'\red_\chi)_x=\A'\red_{\chi(x)}$, where $x$ is any sort or function symbol from $\Sigma$.

\item For all $h':\A'\to\B'\in\Mod(\Sigma')$, $(h'\red_\chi)_s=h'_{\chi(s)}$, where $s$ is any sort from $\Sigma$. 
\end{itemize}
If $\chi$ is an inclusion then we may write $\A'\red_\Sigma$ instead of $\A'\red_\chi$.
For example, if $\chi:\Sigma_\TRIV\to\Sigma_\NAT$ is the renaming described in Example~\ref{ex:triv-nat} and 
$\mathcal{N}$ is the model of natural numbers with addition then 
$\mathcal{N} \red_\chi$ is the set of natural numbers.
\begin{example} 
Let $\mathcal{L}$ be the algebra over $\Sigma_{\mathtt{NATL}}$ defined by:
\begin{itemize}
\item $\mathcal{L}$ interprets all symbols in $\Sigma_\NAT$ as $\mathcal{N}$, the model of natural numbers with addition,

\item $\mathcal{L}_{\mathit{List}}$ consists of all lists of natural numbers, 
while $\mathcal{L}_{\mathit{NeList}}$ consists of all non-empty lists of natural numbers,

\item $\mathcal{L}_{\mathit{empty}}$ is the empty list, 
$\mathcal{L}_{;}$ is the concatenation of lists,
$\mathcal{L}_{\mathit{head}}$ extracts the top of a non-empty list, and
$\mathcal{L}_{\mathit{tail}}$ deletes the first element of a non-empty list.
\end{itemize}
\end{example}
Notice that the reduct of $\mathcal{L}$ to the signature $\Sigma_\NAT$ is $\mathcal{N}$, in symbols, $\mathcal{L}\red_{\Sigma_\NAT}=\mathcal{N}$.
An important result in algebraic specifications, which is at the core of many important developments concerning modularity is the satisfaction condition.
\begin{proposition}[Satisfaction condition]
For all signature morphisms $\chi:\Sigma\to \Sigma'$, all $\Sigma$-sentences $\gamma$ and all order-sorted $\Sigma'$-algebras $\A'$, 
we have  $\A'\models\chi(\gamma)$ iff $\A'\red_\chi\models \gamma$, 
where $\chi(\gamma)$ denotes the translation of $\gamma$ along $\chi$. 
\end{proposition}
Pairs $(\Sigma,E)$ consisting of a signature $\Sigma$ and a set of sentences $E$ over $\Sigma$,
are called presentations in algebraic specification literature.
The notion of signature morphism extends straightforwardly to presentations.
\begin{definition}
A presentation morphism $\chi:(\Sigma,E)\to (\Sigma',E')$ is a signature morphism $\chi:\Sigma\to\Sigma'$ s.t. $E'\models\chi(E)$.
\end{definition}
An example of presentation morphism is the inclusion $\iota:(\Sigma_\NAT,E_\NAT)\hookrightarrow (\Sigma_\mathtt{NATL},E_\mathtt{NATL})$, where $E_\NAT=\{\Forall {x:\mathit{Nat}} 0 + x = x,~\Forall{x,y:\mathit{Nat}}s~y + x = s(x + y)\}$ defined in Example~\ref{ex:eq-nat} and $E_\mathtt{NATL}= E_\NAT \cup \{\Forall{e:\mathit{Elt},l:\mathit{List}}\mathit{head}(e;l)=e, \Forall{e:\mathit{Elt},l:\mathit{List}}\mathit{tail}(e;l)=l\}$.

\paragraph{\bf Proof calculus}

The satisfaction relation $E\models E'$ between sets of sentences is the semantic way to establish truth.
The syntactic approach to truth consists of defining \emph{entailment relations} between sets of sentences involving only syntactic entities.
The correctness of entailment relations can be established only in the presence of satisfaction relation.

\begin{definition}
An entailment system consists of a family of entailment relations $\{\texttt{\_}\!\vdash_\Sigma\! \texttt{\_}\mid \Sigma \text{ is an order-sorted signature}\}$ between sets of sentences with the following properties:
\begin{center}
\begin{tabular}{l l}
$(\mathit{Monotonicity})~\proofrule{}{E\vdash_\Sigma E_1}~[E_1\subseteq E]$ & 

$(\mathit{Union})~\proofrule{E\vdash_\Sigma E_1 \Space E\vdash_\Sigma E_2}{E\vdash_\Sigma E_1\cup E_2}$ \\

& \\

$(\mathit{Translation})~\proofrule{E\vdash_\Sigma E_1\Space}{\chi(E)\vdash_{\Sigma'}\chi(E_1)}~[\chi:\Sigma\to \Sigma']$ &
$(\mathit{Cut})~\proofrule{E \vdash_\Sigma E_1 \Space E\cup E_1\vdash_\Sigma E_2}{E\vdash_\Sigma E_2}$
\end{tabular}
\end{center}
\end{definition}
Notice that we have placed the side conditions of the entailment properties in square brackets.
For example, in case of $(\mi{Monotonicity})$, $E_1\subseteq E$ implies $E\vdash_\Sigma E_1$.
Entailment relations are inductively defined by proof rules.
\begin{definition}[Order-sorted entailment system]\label{def:es}
The entailment system of order-sorted algebra is the least entailment system generated by the following proof rules:
\begin{center}
\begin{tabular}{l l}
$(\mathit{Reflexivity})~\proofrule{}{E\vdash_\Sigma\Forall{X}t=t}$ \Space
$(\mathit{Symmetry})~\proofrule{E\vdash_\Sigma \Forall{X}t_1 =t_2}{E\vdash_\Sigma \Forall{X}t_2 =t_1}$ & 
$(\mathit{Transitivity})~\proofrule{E\vdash_\Sigma \Forall{X}t_1 =t_2\Space E\vdash_\Sigma \Forall{X}t_2 =t_3 }{E\vdash_\Sigma \Forall{X}t_1 =t_3}$ \\
& \\
$(\mathit{Congruence})~\proofrule{E\vdash_\Sigma \Forall{X}t_1 =t_1' ~\dots~ E\vdash_\Sigma \Forall{X}t_n =t_n'}{E\vdash_\Sigma \Forall{X}\sigma(t_1,\dots,t_n) =\sigma(t_1',\dots,t_n')}$ & 
$(\mathit{Substitutivity})~\proofrule{E\vdash_\Sigma \Forall{X} e}{E\vdash_\Sigma\Forall{Y}\theta(e)}~[\theta:X\to T_\Sigma(Y)]$ \\
& \\
$(\mathit{Implication_1})~\proofrule{E\vdash_{\Sigma} t=t' \text{ if }\bigwedge_{i=1}^n t_i=t_i'}{E\cup\{t_1=t_1',\dots, t_n=t_n'\} \vdash_\Sigma t = t'}$ & 
$(\mathit{Implication_2})~\proofrule{E\cup\{t_1=t_1',\dots, t_n=t_n'\} \vdash_\Sigma t = t'}{E\vdash_{\Sigma} t=t' \text{ if }\bigwedge_{i=1}^n t_i=t_i'}$\\
&\\
$(\mathit{Quantification_1})~\proofrule{E\vdash_{\Sigma(X)} e }{E\vdash_\Sigma\Forall{X}e}$ & 
$(\mathit{Quantification_2})~\proofrule{E\vdash_\Sigma\Forall{X}e}{E\vdash_{\Sigma(X)} e}$
\end{tabular}
\end{center}
\end{definition}

\begin{theorem}[Completeness]
The order-sorted equational logic calculus is sound (i.e., $\vdash\subseteq \models$) and complete (i.e. $\models\subseteq \vdash$).
\end{theorem}
Completeness of order-sorted equation deduction is one of the fundamental results in algebraic specification, which has a long tradition starting with Birkhoff \cite{bir-com}, continuing with completeness of many-sorted equational deduction~\cite{gog-com} and order-sorted equational deduction~\cite{orderSorted}, and ending with completeness of institutions with Horn clauses \cite{ctorsLogic}.

An algebraic specification language, in general, and CafeOBJ, in particular, should be understood on  three interconnected semantic levels:
(i) denotational semantics, 
(ii) proof-theoretical definition, and
(ii) operational semantics. 
The denotational semantics consists of classes of models described by the specifications and the satisfaction relation determined by them.
The proof-theoretical aspect is given by the entailment relations, which provides a practical way of reasoning formally about the properties of models described by specifications.
The operational semantics consists of order-sorted term-rewriting~\cite{gog-tpa}, which has several benefits such as an increased level of automation for proofs.

The design of the specification and verification methodology is largely based on the denotational semantics. Therefore, this section focused on describing the underlying logic, order-sorted algebra, and the language constructs such as free semantics, parameterization and imports. The operational semantics of algebraic specification languages is important during formal proofs for evaluating functions by providing the structured, rule-based process to compute values from functions applied to specific arguments. This ensures that every step in the proof is grounded in the formal definitions of the system, leading to correct and an increased automation level in comparison to theorem proving methods which are not powered by term rewriting.

\section{Structured specifications}\label{sec:spec}

The foundational work on module algebra began with the seminal contributions of Professor Jan Bergstra and his collaborators~\cite{BergstraModuleAlgebra}. To the best of our knowledge, this marked the start of the concept of module algebra and the formal study of rules for constructing module expressions—referred to here as structured specifications—which are used to describe software systems.\footnote{To be more precise, in this context a \emph{module} is a specification with a name.} Rather than reiterating the well-established importance of modularizing software, we emphasize that the mathematical foundations supporting this modularization are grounded in the specification-building operators that define structured specifications. These operators provide the formal mechanisms for composing and manipulating modular software components.
We refer the reader to~\cite{DiaconescuT11,DiaconescuTutu14,Tutu14,SanellaTarkecki12} for the latest developments in this area.
This section is devoted to a brief presentation of fundamentals of structured specifications and their application to the specification methodology.

\subsection{Reachability}
 
In practice, only order-sorted algebras that are \emph{reachable} by some constructor operations are of interest. 
Consider, for example, the signature of natural numbers with addition $\NAT$ described in Example~\ref{ex:sig-nat}.
One model of interest is $\mathcal{N}$, the model of natural numbers.
Other model of interest can be $\mathcal{Z}_n$, the model of integers modulo $n$, where $n>1$, which interprets both sorts $\mi{Nat}$ and $\mi{NzNat}$ as $\{\widehat{0},\widehat{1},\dots,\widehat{n-1}\}$ and the function symbols in the usual way.
Both $\mathcal{N}$ and $\mathcal{Z}_n$ are reachable by the operations $0:~\to \mi{Nat}$ and $s\_:\mi{Nat}\to \mi{NzNat}$, that is, the unique homomorphisms $\mathcal{T}_\NAT\to \mathcal{N}$ and $\mathcal{T}_\NAT\to \mathcal{Z}_n$ are surjective.
Next, we present constrained and loose sorts, and then
we generalize the concept of reachability to any signature $\Sigma=(S,\leq,F)$ for which we distinguish a subset $F^c\subseteq F$ of constructors.

\begin{definition}[Constrained and loose sorts]
A sort $s\in S$ is called \emph{constrained} if $s$ has a constructor $\sigma:w\to s\in F^c$, 
otherwise $s$ is called loose. 
\end{definition}

\begin{definition}[Reachable algebras]
An order-sorted algebra $\A$ over a signature $\Sigma=(S,\leq,F)$ is reachable by some constructors $F^c\subseteq F$ if there exists a set of $X$ of variables of loose sort and a valuation $f:X\to \A$ such that the unique extension $f^\#:\mathcal{T}_{\Sigma^c}(X)\to \A\red_{\Sigma^c}$ of $f:X\to \A$ to a $\Sigma^c$-homomorphism is surjective, where $\Sigma^c=(S,\leq,F^c)$. 
\end{definition} 

We give some examples of reachable algebras with loose sorts.
\begin{example} \label{ex:sig-bag}
Let $\Sigma_\mathtt{BAG}$ be the signature defined as follows:
\begin{itemize}
\item The set of sorts is $\{\mi{Elt}, \mi{Bag}\}$ such that $\mi{Elt} \leq \mi{Bag}$. 
\item The set of function symbols is $\{\mi{empty} : ~ \to \mi{Bag}, \_\circ\_ : \mi{Bag} ~ \mi{Bag} \to \mi{Bag} , \mi{take} : \mi{Bag} ~ \mi{Elt} \to \mi{Bag} \}$.
\end{itemize}
\end{example}
Let $F_\mathtt{BAG}^c=\{\mi{empty} : ~ \to \mi{Bag}, \_\circ\_ : \mi{Bag} ~ \mi{Bag} \to \mi{Bag}\}$ be a set of constructors for the signature $\Sigma_\mathtt{BAG}$.
The sort $\mi{Elt}$ is loose while the sort $\mi{Bag}$ is constrained.
We give some examples of reachable algebras by the constructors in $F_\mathtt{BAG}^c$:

\begin{enumerate}
\item the model of sets of natural numbers, denoted $\B_1$, which interprets $\mi{empty}$ as the empty set, $\circ$ as the union of sets, and $\mi{take}$ as the extraction of an element from a set.

\item the model of strings with elements from $\{a,b\}$, denoted $\B_2$, which interprets $\mi{empty}$ as the empty string $\circ$ as the concatenation of strings and $\mi{take}$ as the extraction of the first occurrence of an element in a string.

\item the model of integers, denoted $\B_3$, which interprets both sorts $\mi{Elt}$ and $\mi{Bag}$ as the set of integers, $\mi{empty}$ as zero, $\circ$ as addition and $\mi{take}$ as subtraction.
\end{enumerate}
For the first case, let $X=\{x_i\mid i\in \N \}$ be a countably infinite set of variables, and $f:X\to \B_1$ a valuation defined by $f(x_i)=n$ for all $n\in\N$.
It is easy to check that the extension $f^\#:T_{\Sigma_\mathtt{BAG}^c}(X)\to \B_1\red_{\Sigma_\mathtt{BAG}^c}$ of $f:X\to \B_1$ to a $\Sigma_\mathtt{BAG}^c$-homomorphism is surjective.
Similar arguments can be used to show that $\B_2$ and $\B_3$ are reachable by $F_\mathtt{BAG}^c$.

\subsection{Congruences}
Given a signature $\Sigma=(S,\leq ,F)$, a \emph{congruence} $\equiv$ on an algebra $\A$ is an $S$-sorted equivalence on $\A$ (that is, an equivalence $\equiv_s$ on $\A_s$ for all sorts $s\in S$) compatible with
\begin{itemize}
\item the functions, that is, if  $a_i\equiv_{s_i} b_i$ for all $i\in\{1,\dots,n\}$ then $\A_\sigma(a_1,\dots,a_n)\equiv_s\A_\sigma(b_1\dots,b_n)$, for all function symbols $\sigma:s_1\dots s_n\to s\in F$, and

\item the sort order, that is, $a \equiv_{s_1} b$ iff $a \equiv_{s_2} b$ for all sorts  $s_1,s_2\in S$ such that $s_1\equiv_\leq s_2$ and all elements $a,b\in \A_{s_1}\cap\A_{s_2}$.
\end{itemize}
Any set of conditional equations $E$ generates a congruence on the initial model of terms $\mathcal{T}_\Sigma$ as follows:
$$\equiv^E\coloneqq \{(t, t')\mid E\models t=t' \}$$ where $E\models t=t'$ means that $\A\models E$ implies $\A\models t=t'$ for all $\Sigma$-algebras $\A$. 
The quotient algebra $\mathcal{T}_\Sigma/_{\equiv^E}$ denoted $\mathcal{T}_{(\Sigma,E)}$ has the following initial property~\cite{orderSorted}:
for each algebra $A$ satisfying $E$, there exists a unique homomorphism $\mathcal{T}_{(\Sigma,E)}\to \A$.
The initial algebra $\mathcal{T}_{(\Sigma,E)}$ of $E$ is unique up to isomorphism and it gives the \emph{tight denotation} of the equational specification $(\Sigma,E)$.

Recall the signature of natural numbers with addition defined in Example~\ref{ex:sig-nat} and the set of sentences $E_\NAT$ of Example~\ref{ex:eq-nat}.
Notice that $\mathcal{N}$, the model of natural numbers that interprets all function symbols in the usual way, is isomorphic to $T_{(\Sigma_\NAT,E_\NAT)}$ the initial algebra of $E_\NAT$.

\subsection{Free models}
In algebraic specification, the notion of free algebra is one of the basic concepts, which is a generalization of the initial algebra. 
In initial semantics, the idea is to define a minimal model of the specification where the behavior of the data type and its operations are determined uniquely.
The initial algebra is important because it guarantees that if two terms are equal under the specified operations (according to the equations), then they are equal in the free algebra. This forms the basis of equational reasoning in algebraic specifications.
When using initial semantics, the free algebra acts as the default or ``best'' interpretation of the abstract data type, ensuring that no unintended properties or relationships are introduced.
In addition, free models cannot be fully specified by first-order theories because the set of semantic consequences of a first-order theory is recursively enumerable, whereas the properties of free models---when formalized as sentences---are not, due to G\"odel's incompleteness theorem. This theorem implies that there are true statements about free models that cannot be derived from any first-order theory, meaning such models may possess properties that lie beyond the expressive power of first-order logic to capture in a complete and consistent manner.

An example of free algebra is given by the $\mathtt{NATL}$-model of lists with elements from $\mathcal{Z}_2$, the algebra of integers modulo $2$ which interprets the symbols from $\Sigma_\NAT$ in the usual way.
In the following, we will describe briefly its construction.

The notion of free algebra is best explained using signature morphisms.
Let $\chi:\Sigma\to\Sigma'$ be a signature morphism and $\A$ a $\Sigma$-algebra.
Without loss of generality, we assume that the elements of $\A$ are different from the constants of both $\Sigma$ and $\Sigma'$.
In this case, each element $a\in \A_s$ can be regarded as a new constant $a:~\to s$ for $\Sigma$, where $s$ is any sort of~$\Sigma$.
Let $\Sigma_\A$ be the signature obtained from $\Sigma$ by adding the elements of $\A$ as new constants.
Similarly, we let $\Sigma'_\A$ denote the signature obtained from $\Sigma'$ by adding elements $a\in \A_s$ as new constants $a:~\to\chi(s)$, where $s$ is any sort of~$\Sigma$.
The result is a new signature morphism $\chi_\A:\Sigma_\A\to\Sigma'_\A$ which extends $\chi$ by mapping each constant $a:~\to s$ to $a:~\to\chi(s)$, for all sorts $s$ of $\Sigma$ and all elements $a\in\A_s$.
\begin{center}
\begin{tikzcd}[row sep=large, column sep=large]
\mathcal{T}_{(\Sigma',E')}(\A)\models E' \arrow[r,dash,dotted]& \Sigma' \arrow[r,hook,dashed]& \Sigma'_\A & \arrow[l,dash,dotted] \mathcal{T}_{(\Sigma'_\A,E'\cup E'_\A)}\models E'\cup E'_\A \arrow[lll,swap,"\red_{\Sigma'}",bend right=7,dotted,end anchor={[shift={(30pt,0pt)}]north west}] \\
\A \arrow[r,dash,dotted] & \Sigma \arrow [u,"\chi"] \arrow[r,hook,dashed] & \Sigma_\A \arrow[u,swap,"\chi_\A"] & \arrow[l,dotted,dash] \A_\A\models E_\A \arrow[lll, "\red_\Sigma", bend left=10,dotted]
\end{tikzcd}
\end{center}
We let $\A_\A$ denote the expansion of $\A$ to $\Sigma_\A$, which interprets each constant $a:~\to s$ as $a$, for all sorts $s$ of $\Sigma$ and all elements $a\in\A_s$.
We denote by $E_\A$ the set of all ground $\Sigma_\A$-equations satisfied by $\A_\A$, and by $E'_\A$ the translation of $E_\A$ along $\chi_\A$.
Notice that $\A_\A$ is the initial algebra of $E_\A$.
For any set of conditional $\Sigma'$-equations $E'$, 
the \emph{free $(\Sigma',E')$-algebra over $\A$ via $\chi$}, 
denoted $\mathcal{T}_{(\Sigma',E')}(\A)$,
is $\mathcal{T}_{(\Sigma'_\A,E'\cup E'_\A)}\red_{\Sigma'}$, 
where $\mathcal{T}_{(\Sigma'_\A,E'\cup E'_\A)}$ is the initial algebra of $E'\cup E'_\A$.
Since $\A_\A$ is the initial algebra of $E_\A$ and $\mathcal{T}_{(\Sigma'_\A,E'\cup E'_\A)}\models E'_\A$,
there exists a unique arrow $\eta_\A:\A_\A\to \mathcal{T}_{(\Sigma'_\A,E'\cup E'_\A)}\red_{\chi_\A}$.
The reduct $\eta:\A\to \mathcal{T}_{(\Sigma',E')}(\A)\red_\chi$ of 
$\eta_\A:\A_\A\to \mathcal{T}_{(\Sigma'_\A,E'\cup E'_\A)}\red_{\chi_\A}$ to the signature $\Sigma$ 
is a universal arrow to the functor $\red_\chi\_:\Mod(\Sigma',E')\to\Mod(\Sigma)$, that is,
for each $\Sigma'$-algebra $\B'$ that satisfies $E'$ and 
any homomorphism $h:\A\to \B'\red_\chi$ 
there exists a unique $\Sigma'$-homomorphism $h':\mathcal{T}_{(\Sigma',E')}(\A)\to \B'$ such that $\eta;h'\red_\chi=h$.\footnote{$\Mod(\Sigma',E')$ is the full subcategory of $\Mod(\Sigma')$ of all $\Sigma'$-algebras satisfying $E'$.}
\begin{center}
\begin{tikzcd}[row sep=large, column sep=large]
 \A \arrow[r,"\eta"] \arrow[dr,swap,"h"] & 
 \mathcal{T}_{(\Sigma',E')}(\A)\red_\chi \arrow[d,"h'\red_\chi",dashed] &  
 \mathcal{T}_{(\Sigma',E')}(\A) \arrow[d,dashed,"\exists ! ~ h'"]\\
 & \B'\red_\chi & \B'
\end{tikzcd}
\end{center}
Consider the signature inclusion $\Sigma_\NAT\hookrightarrow \Sigma_\mathtt{NATL}$, defined in Example~\ref{ex:nat-list}, and $\mathcal{Z}_2$, the $\Sigma_\NAT$-algebra of integers modulo~2.
Let $E_{\mathtt{NATL}}\coloneqq E_\NAT \cup \{\Forall{e:\mi{Elt},l:\mi{List}}\mi{head}(e;l)=e, \Forall{e:\mi{Elt},l:\mi{List}}\mi{tail}(e;l)=l\}$, where $E_\NAT$ is the set of sentences defined in Example~\ref{ex:eq-nat}.
The model of lists with elements from $\mathcal{Z}_2$ is formally defined as the free $(\Sigma_{\mathtt{NATL}},E_{\mathtt{NATL}})$-algebra over $\mathcal{Z}_2$, that is, $\mathcal{T}_{(\Sigma_\mathtt{NATL},E_\mathtt{NATL})}(\mathcal{Z}_2)$.
In this case, the universal arrow $\eta:\mathcal{Z}_2\to \mathcal{T}_{(\Sigma_\mathtt{NATL},E_\mathtt{NATL})}(\mathcal{Z}_2)\red_{\Sigma_\NAT}$ is the identity.

\subsection{ Specification building operators}
Structured specifications are constructed from some basic specifications by iteration of several specification building operators.
In algebraic specification languages such as CafeOBJ or Maude (see Section~\ref{sec:tools}), the definition of all language constructs is based on specification building operators, which can be regarded as primitive operators.
In software engineering, structuring constructs give the possibility of systematic reuse of already defined \emph{modules}, which are in fact  labeled specifications.
The semantics of a specification $\SP$ consists of a signature $\Sigma_\SP$ and a class of models $\M_\SP$. 
For any structured specification $\SP$, we can also define its set of sentences $E_\SP$.
\begin{itemize}
\item [\bf BASIC]
A basic specification is a presentation $(\Sigma,E)$, where $\Sigma$ is a signature and $E$ is a set of sentences over $\Sigma$.

$\Sigma_{(\Sigma,E)}=\Sigma$, $E_{(\Sigma,E)}=E$, and $\M_{(\Sigma,E)}=\{\A\in|\Mod(\Sigma)| \mid \A\models E \}$.

\item [\bf CONS]
For any specification $\SP$, the restriction of $\SP$ to some constructor operators $F^c$ from $\Sigma_\SP$, denoted $\SP|_{F^c}$, is defined by
$\Sigma_{(\SP|_{F^c})}=\Sigma_\SP$, 
$E_{(\SP|_{F^c})}=E_\SP$, and 
$\M_{(\SP|_{F^c})}=\{\A\in \M_\SP \mid \A \text{ is reachable by } F^c\}$.

\item [\bf UNION] 
For any specifications $\SP_1$ and $\SP_2$ with the same signature $\Sigma$, the union $\SP_1\cup\SP_2$ is defined by

$\Sigma_{(\SP_1\cup\SP_2)}=\Sigma$, 
$E_{(\SP_1\cup\SP_2)}=E_{\SP_1}\cup E_{\SP_2}$, and 
$\M_{(\SP_1\cup\SP_2)}=\M_{\SP_1}\cap \M_{\SP_2}$.

\item[\bf TRANS] 
The translation of $\SP$ along a signature morphism $\chi:\Sig(\SP)\to \Sigma$ denoted by $\SP\star \chi $ is defined by

$\Sigma_{(\SP\star\chi)}=\Sigma$, $E_{(\SP\star\chi)}=\chi(E_\SP)$, and 
$\M_{(\SP\star\chi)}=\{\A \in|\Mod(\Sigma)| \mid \A\red_\chi\in \M_\SP \}$.

\item [\bf $\mathcal{H}$-FREE]
Given a class of homomorphisms $\mathcal{H}$, 
for all specifications $\SP$, 
all basic specifications $(\Sigma,E)$, and 
all presentation morphisms $\chi:(\Sigma_\SP,E_\SP)\to (\Sigma,E)$, 
the $\mathcal{H}$-free $(\Sigma,E)$-specification over $\SP$ via $\chi$,
denoted $(\Sigma,E)!_{\mathcal{H}}\SP$, is defined~by

$\Sigma_{(\Sigma,E)!_{\mathcal{H}}\SP}=\Sigma$,
$E_{(\Sigma,E)!_{\mathcal{H}}\SP}=E$, and

$\M_{(\Sigma,E)!_{\mathcal{H}}\SP}= \{T_{(\Sigma,E)}(\A) \mid  \A\in\M_\SP \text{ with the universal arrow } \eta:\A\to T_{(\Sigma,E)}(\A)\red_\chi \text{ in } \mathcal{H} \}$.
\end{itemize}
$\mathcal{H}$-\textbf{FREE} is the standard free semantics operator.
This operator is parameterized by a class of homomorphisms $\mathcal{H}$, which is one of the following ones:
(a)~$\ID$, the class of identities, 
(b)~$\EX$, the class of inclusions, or 
(c)~$\US$, the class of all homomorphisms.
Identities yield \emph{protecting} imports, which do neither collapse elements nor add new elements to the models of the imported module. 
In the algebraic specification literature, these conditions are known as ``no junk and no confusion'' conditions, respectively.
Inclusions yield importations that allow ``junk'' but forbid ``confusion.''
If $\mathcal{H}$ consists of all homomorphisms then both ``junk'' and ``confusion'' are allowed.

There are other specification building operators defined in the algebraic specification literature, which are useful in practice and cannot be derived from the ones defined in this survey. 
Some examples are the derivation across signature morphisms used for hiding information \cite{SannellaT88} or the extension operator used for capturing non-protecting importation modes~\cite{DiaconescuT11}.
A basic property of structured specifications is given in the following lemma.
\begin{lemma}
For each structured specification $\SP$, $\A\in\M_\SP$ implies $\A\models E_\SP$.
\end{lemma}

The predefined module $\BOOL$ is, perhaps, the most used (labelled) specification in the OBJ languages.
$\BOOL$ is not a specification of Boolean algebras.
It defines the truth values used in the verification methodology.
\begin{example}[Booleans] \label{ex:bool}
The following is the specification of truth values $\mi{true}$ and $\mi{false}$ with the usual operations defined on them.
\codesize\begin{verbatim}
   mod! BOOL{
    [Bool]                              eq true and A = A .                                           
    op true : -> Bool                   eq false and A = false .                         
    op false : -> Bool                  eq A and A = A . 
    op _and_ : Bool Bool -> Bool        eq false xor A = A . 
    op _or_ : Bool Bool -> Bool         eq A xor A = false .
    op _xor_ : Bool Bool -> Bool        eq A and (B xor C) = A and B xor A and C .
    op not_ : Bool -> Bool              eq not A = A xor true .
    op _implies_ : Bool Bool -> Bool    eq A or B = A and B xor A xor B .
    vars A B C : Bool .                 eq A implies B = not (A xor A and B) . }           
\end{verbatim}
\end{example}
$\BOOL$ is defined with initial semantics, which means that the denotational semantics of $\BOOL$ consists of (the class of all algebras isomorphic to) the initial algebra of $(\Sigma_\BOOL,E_\BOOL)$, that is, the algebra with the carrier set $\{\mi{true},\mi{false}\}$ that interprets the Boolean operations in the usual way.
The free semantics has many implications in both specification and verification methodologies, which will be explained gradually in the present contribution.

\begin{example}[Labels] \label{ex:label}
The following specification defines some labels denoting the state of programs/processes trying to access a common resource.
\codesize\begin{verbatim}
   mod* LABEL{
    pr(BOOL)                           eq [e1]: (L = L) = true .
    sort Label                         ceq[e2]: L1 = L2 if (L1 = L2) = true . 
    ops rs ws cs : -> Label {constr}   eq [l1]: (rs = cs) = false .
    op _=_ : Label Label -> Bool       eq [l2]: (ws = cs) = false .
    vars L L1 L2 : Label               eq [l3]: (cs = rs) = false . }
\end{verbatim}
\end{example}

The specification $\BOOL$ is imported in protecting mode.
Therefore, $\Sigma_\LABEL$ is obtained from $\Sigma_\BOOL$ by adding the symbols described on the left column of Example~\ref{ex:label}.
The constants $\mathit{rs}$, $\mathit{ws}$, and $\mathit{cs}$ denote the reminder section, waiting section, and critical section, respectively.
The operation $\texttt{\_=\_} : \Label ~\Label \to \Bool$ denotes the equality.
By the first two equations, $e1$ and $e2$, the object level equality denotes the real equality among labels, that is,
two labels $a,b$ are equal in a model $\A\in\M_\LABEL$ iff $\A_=(a,b)=\A_{true}$.
Since $\BOOL$ is imported in protecting mode, the equations $l1$ --- $l3$ say that no ``confusion'' is allowed between the labels 
$\mathit{rs}$, $\mathit{ws}$, and~$\mathit{cs}$. Because the constants $\mathit{rs}$, $\mathit{ws}$, and~$\mathit{cs}$ are also
constructors, no ``junk'' is allowed to the sort $\Label$.
It follows that the semantics of $\LABEL$ consists of the initial model.

Note that the importation of $\BOOL$ in protecting mode and the equations $e1,e2$ allow one to write sentences semantically equivalent to inequalities. 
For example, $l1$, $l2$, and $l3$ are semantically equivalent to $rs \neq cs$, $ws\neq cs$, and $cs \neq rs$, respectively.
Unsurprisingly, there are structure specifications which are inconsistent, i.e., with no models.

Since $\LABEL$ imports $\BOOL$ in protecting mode, there exists a signature inclusion $\iota:\Sigma_\BOOL\hookrightarrow \Sigma_\LABEL$.
If we ignore the constructor declarations, the left column of Example~\ref{ex:label} corresponds to $\BOOL\star\iota$, while the right column to $(\Sigma_\LABEL,\{e1,e2,l1,l2,l3\})$.
Let $F_\LABEL^c$ denote the set of constructors 
$\{\mathit{rs}:\to\Label,\mathit{ws}:\to\Label,\mathit{cs}:\to\Label\}$ and notice that $\LABEL=(\BOOL\star\iota \cup (\Sigma_\LABEL,E_\BOOL \cup \{e1,e2,l1,l2,l3\}))|_{F_\LABEL^c}$. 

\begin{example}\label{ex:omega}
The following is a specification of the usual ordering on natural numbers, which is then extended to an ordering on $\omega + 1$.
\codesize\begin{verbatim}
 mod! PNAT{                                 
  protecting(BOOL)                             
  [Nat]                                        
  op 0 : -> Nat                               mod! OMEGA{ 
  op s_ : Nat -> Nat                           extending(PNAT)
  op _<=_ : Nat Nat -> Bool                    op omega : -> Nat
  vars X Y : Nat                               vars X Y : Nat 
  eq [o1]: X <= X = true .                     eq [o3]: s omega = omega .
  cq [o2]: X <= s Y = true if X <= Y .}        eq [o4]: X <= omega = true .}
\end{verbatim}
\end{example}
Since $\BOOL$ is imported in protecting mode, $\Sigma_\PNAT$ is obtained from $\Sigma_\BOOL$ by adding the sort $\Nat$ and the function symbols described on the left column of Example~\ref{ex:omega}.
We have $\PNAT= (\Sigma_\PNAT,E_\BOOL \cup \{o1,o2\}) !_\ID \BOOL$. 
Similarly, $\Sigma_\OMEGA$ is obtained from $\Sigma_\PNAT$ by adding the constant $\mi{omega}:\to \mi{Nat}$.
It follows that $\OMEGA=(\Sigma_\OMEGA,E_\PNAT \cup \{o3,o4\}) !_\EX \PNAT$.

\subsection{Parameterized specifications}
Another important concept for the modularization of system specifications is that of parameterized specification, which plays a crucial role in scaling up to the complexity of software.
The basic mechanism for the instantiation of parameters relies on pushouts of signature morphisms.
The concept of parameterized specification and of pushout-style instantiation of parameters originates from the work on Clear~\cite{BurstallG77IJCAI} and constitutes the basis of parameterized specifications in OBJ family of languages.

Presentation morphisms are extended naturally to specifications.
A \emph{specification morphism} $\chi:\SP\to\SP'$ is a signature morphism $\chi:\Sigma_\SP \to \Sigma_{\SP'}$ such that for all $\A'\in\M_{\SP'}$ we have $\A\red_\chi\in\M_\SP$.
If both $\SP$ and $\SP'$ are structured specifications then we require also that $E_{\SP'}\models \chi(E_\SP)$.
\begin{definition}[Parameterized specification] \label{def:parameter}
 A parameterized specification $\SP(\mathtt{P})$ is a specification morphism $\mathtt{P}\to \SP$ such that the underlying signature morphism is an inclusion $\Sigma_\mathtt{P} \subseteq \Sigma_\SP$ which 
\begin{enumerate}
\item does not allow new subsorts for $\mathtt{P}$, that is, for all $s\in S_\SP$ and all $s_0\in S_\mathtt{P}$ such that $s \leq_\SP s_0$, we have $s\leq_\mathtt{P} s_0$, and 
\item preserves the subsort polymorphic families of operations, that is, for all $\sigma:w \to s \in F_\mathtt{P}$ and all $\sigma:w'\to s'\in F_\SP$ such that $w\equiv_{\leq_\SP} w'$, we have $\sigma:w'\to s'\in F_\mathtt{P}$.
\end{enumerate}
The specification $\mathtt{P}$ is the \emph{parameter} and the specification $\SP$ is the \emph{body} of the parameterized specification $\SP(\mathtt{P})$.
\end{definition}
The conditions of Definition~\ref{def:parameter} will ensure that the instantiation of parameters is well-defined~\cite{GainaNOF20}.
From a practical point of view, parameters are importations in protecting mode.
We give an example of parameterized specification written in CafeOBJ notation.
\begin{example}[Queue] \label{ex:queue}
Let $\TRIV$ be the specification that consists of only one sort $\mi{Elt}$, that is, $\Sigma_\TRIV=(\{\mi{Elt}\},\emptyset,\emptyset)$ and $E_\TRIV=\emptyset$.
The following is a specification of queues of arbitrary elements:
\begin{verbatim}
  mod! QUEUE (X :: TRIV) {                       
   [EQueue NeQueue < Queue]                     var Q : Queue 
   op empty : -> EQueue {constr}                var E F : Elt.X 
   op _;_ : Elt.X Queue -> NeQueue {constr}     eq [q1]: top(E ; Q) = E . 
   op head : NeQueue -> Elt.X                   eq [q2]: head(E ; Q)= Q .
   op tail : NeQueue -> Queue                   eq [q3]: add(empty,F) = F ; empty .  
   op add : Queue Elt.X -> Queue                eq [q4]: add(E ; Q, F)= E ; add(Q,F). }
\end{verbatim}
\end{example}
The parameter of $\QUEUE$ is $\TRIV$.
The signature $\Sigma_\QUEUE$ is obtained from $\Sigma_\TRIV=(\{\mi{Elt}\},\emptyset,\emptyset)$ by adding the function symbols defined on the right column of Example~\ref{ex:queue}, and $E_\QUEUE=\{q1,q2,q3,q4\}$.
Obviously, the inclusion $\Sigma_\TRIV\hookrightarrow \Sigma_\QUEUE$ satisfies the reflection and preservation conditions required by Definition~\ref{def:parameter}.
Since $\QUEUE$ is defined with free semantics and $\TRIV$ is imported in protecting mode,
$\QUEUE= (\Sigma_\QUEUE,E_\QUEUE)!_\ID \TRIV$.

Recall the specification of natural numbers of Example~\ref{ex:omega}, and assume we want a specification of lists of natural numbers.
Let $v:\TRIV\to \PNAT$ be the specification morphism which maps $\mi{Elt}$ to $\mi{Nat}$.
The instance of $\QUEUE(\TRIV)$ by $v$, denoted $\QUEUE(\TRIV\Leftarrow v)$, or simply, $\QUEUE(\PNAT)$ if there is no danger of confusion, is defined as the pushout of $\QUEUE\stackrel{\chi}\hookleftarrow \TRIV\stackrel{v}\to \PNAT$:
\begin{enumerate}
\item The signature of $\Sigma_{\QUEUE(\PNAT)}$ is the vertex of the following pushout of signatures.
\begin{center}
\begin{tikzcd}
\Sigma_{\QUEUE(\TRIV)} \ar[r,"v'",dashed]& \Sigma_{\QUEUE(\PNAT)} \\
\Sigma_\TRIV \ar[u,"\chi",hook] \ar[r,"v",swap]& \Sigma_{\PNAT} \ar[u,"\chi'",swap,hook,dashed]
\end{tikzcd}
\end{center}
The vertex $\Sigma_{\QUEUE(\PNAT)}$ of the above pushout is obtained from $\Sigma_{\QUEUE(\TRIV)}$ by substituting the sort $\Nat$ for $\Elt$ and by adding all the function symbols from $\Sigma_{\PNAT}$. 

\item Let $\mathcal{N}$ be the unique (up to isomorphism) algebra of $\PNAT$ and let $\N = {\mathcal{N}\red_v}$.
We denote by $\mathcal{L}$ the free $(\Sigma_{\QUEUE(\TRIV)},E_{\QUEUE(\TRIV)})$-algebra over $\N$ and notice that $\mathcal{L}\red_\chi=\N$.
By \cite[Theorem~23]{GainaNOF20}, there exists a unique algebra $\mathcal{L}'$ such that $\mathcal{L}'\red_{v'}=\mathcal{L}$ and $\mathcal{L}'\red_{\chi'}=\mathcal{N}$,
which is exactly the free $(\Sigma_{\QUEUE(\PNAT)},E_{\QUEUE(\PNAT)})$-algebra over $\mathcal{N}$.
The class $\M(\SP)$ consists of all algebras isomorphic to $\mathcal{L}'$.

\item $E_{\QUEUE(\PNAT)} = v'(E_{\QUEUE(\TRIV)})\cup E_{\PNAT}$.
\end{enumerate}
It follows that $\QUEUE(\PNAT)= (\Sigma_{\QUEUE(\PNAT)},E_{\QUEUE(\PNAT)})!_\ID \PNAT$.

\subsection{Specification proof calculus}
In this section, we define an entailment system for reasoning formally about the properties of structured specifications.
Its definition is based on the entailment system described in Definition~\ref{def:es} for inferring new sentences for an initial set of axioms.

\begin{definition}
An entailment system for reasoning formally about the logical consequences of structured specifications consists of a family of unary relations on the set of sentences
$\{\SP ~ {\vdash\_}\mid \SP \text{ is a structured specification}\}$ satisfying the following properties:
\begin{center}
\begin{tabular}{l l}
$(\mi{Monotonicity})~\proofrule{}{\SP\vdash E}~[E\subseteq E_\SP]$ & 
$(\mi{Union})~\proofrule{\SP\vdash E_1 \Space \SP\vdash E_2}{\SP\vdash E_1\cup E_2}$ \\
& \\
$(\mi{Translation})~\proofrule{\SP\vdash E}{\SP\star \chi\vdash \chi(E)}~[\chi:\Sigma_\SP\to \Sigma]$& 
$(\mi{Cut})~\proofrule{\SP \vdash E_1 \Space \SP \cup (\Sigma_\SP,E_1) \vdash E_2}{\SP\vdash E_2}$
\end{tabular}
\end{center}
\end{definition}

The operational semantics of a specification $\SP$ is obtained by orienting the equations in $E_\SP$ from left to right.
An unconditional term rewriting step is defined by 
$C[l\theta]\Rightarrow_\SP C[r\theta]$ 
for all equations $\Forall{X}l=r\in E_\SP$, all substitutions $\theta:X\to T_{(\Sigma_\SP)}(Y)$ and all contexts $C$, that is, terms with exactly one occurrence of a special 
variable $\Box$.\footnote{Notice that $C[l\theta]$ denotes the term obtained from $C$ by substituting the term $l\theta$ for the variable $\Box$.}
The term rewriting relation $\stackrel{*}\Rightarrow_\SP$ is the reflexive-transitive closure of $\Rightarrow_\SP$.
For details about the definition of term rewriting relation, we recommend \cite{TeReSe}.
Two terms $t_1$ and $t_2$ are \emph{joinable}, in symbols, $\SP\vdash t_1\downarrow t_2$  if there exists another term $t$ such that $t_1\stackrel{*}\Rightarrow_\SP t$ and $t_2\stackrel{*}\Rightarrow_\SP t$.
To understand the proof score methodology, it is not necessary to know the details of term rewriting relation, and its implementation can be regarded as a black-box.

A specification $\SP$ satisfies a set of sentences $E$ over $\Sigma_\SP$, in symbols, $\SP\models E$ iff $\A\models E$ for all order-sorted algebras $\A\in \M_\SP$. 
An example of structured entailment system is the semantic structured entailment system $\{\SP\models \_ \mid \SP \text{ is a structured specification}\}$.
To check that $\SP\models E$ one needs to consider all order-sorted algebras $\A\in\M_\SP$, which is not feasible, since $\M_\SP$ is a class (not even a set).  
Therefore, an entailment system generated by a finite set of proof rules is proposed.

\begin{definition}[Order-sorted specification calculus]
The order-sorted specification calculus consists of the least entailment system for structured specifications closed under the following proof rules:
\begin{center}
\begin{tabular}{l l}
$(\mi{Implication_1})~\proofrule{\SP\vdash t_0=t_0'\text{ if }\bigwedge_{i=1}^n t_i=t_i'}{\SP\cup (\Sigma_\SP,\{t_1=t_1,\dots,t_n=t'_n)\vdash t_0=t_0' }$ & 
$(\mi{Implication_2})~\proofrule{\SP\cup (\Sigma_\SP,\{t_1=t_1',\dots, t_n=t_n'\})\vdash t_0=t_0' }{\SP\vdash t_0=t_0'\text{ if }\bigwedge_{i=1}^n t_i=t_i'}$ \\
& \\
$(\mi{Quantification_1})~\proofrule{\SP\vdash \Forall{X}e }{ \SP\star\chi \vdash e} ~[~\chi:\Sigma_\SP\hookrightarrow \Sigma_\SP(X)~]$ & 
$(\mi{Quantification_2})~\proofrule{ \SP\star\chi \vdash e}{\SP\vdash \Forall{X}e } ~[~\chi:\Sigma_\SP\hookrightarrow \Sigma_\SP(X)~]$\\
& \\
$(\mi{Rewriting})~\proofrule{(\Sigma_\SP,E_\SP) \vdash t_1 \downarrow t_2}{\SP\vdash \Forall{X}t_1 = t_2}$ \hspace{0.5cm}
\rlap{$(\mi{Ind})~\proofrule{ \SP\star\chi^\sigma \cup (\Sigma_\SP(X^\sigma),\IH^\sigma ) \vdash  e^\sigma\text{ for all } \sigma:w\to s \in F^c_\SP \text{ with } s\leq_\SP s_x }{ \SP\vdash \Forall{x}e}$}\\
\end{tabular}
\end{center}

\noindent
where $s_x$ is the sort of $x$,
$F^c_\SP$ consists of all the constructors of $\SP$,
$X^\sigma=\{x_1 \dots x_n\}$ is a set of new variables, one for each sort in the arity $w=s_1\dots s_n$,
$\chi^\sigma:\Sigma_\SP\hookrightarrow \Sigma_\SP(X^\sigma)$ is a signature inclusion,
$\IH^\sigma$ is the induction hypothesis $\{e\theta \mid \theta: \{x\} \to X^\sigma \text{ is a sort decreasing mapping} \}$, and
$e^\sigma$ is obtained from $e$ by substituting $\sigma(x_1,\dotsm,x_n)$ for $x$.
\end{definition}
A few remarks are in order.
$(\mi{Reflexivity})$, $(\mi{Symmetry})$, $(\mi{Transitivity})$, $(\mi{Congruence})$ and $(\mi{Substitutivity})$ are replaced by $(\mi{Rewriting})$.
According to \cite{gog-tpa}, we have $E_\SP\vdash \Forall{X} t_1=t_2$ iff $(\Sigma_\SP,E_\SP)\vdash t_1\downarrow t_2$ provided that the term rewriting system $(\Sigma_\SP,E_\SP)$ is \emph{confluent} and \emph{terminating}.
Structural induction $(\mathit{Ind})$ is sound for all reachable models, which are defined by the specification building operator \textbf{CONS} in the presence of some constructor operators. 
\section{The tools}\label{sec:tools}

We present in this section the different implementations where proof scores can be executed.
We discuss their unique features and the main differences between them.
Although a detailed comparison of these programming languages is beyond the scope of this
paper, interested readers can find a simple mutual exclusion protocol for two processes implemented
for all of them at
\url{https://github.com/ariesco/proof-scores-survey}.

\subsection{OBJ3}\label{subsec:obj3}

Although it is currently discontinued, OBJ3~\cite{obj3} was the first programming supporting
verification by proof scores. Because CafeOBJ and Maude are sister languages extending
OBJ3, it is worth discussing its main features.

In OBJ3 modules are called objects, which can be defined with both tight and loose semantics.
Sorts, operators, equational axioms, and equations can be stated in objects. The predefined
equality between terms checks syntactic equality,
which might be inconvenient when a term cannot be completely reduced (it would
return \texttt{false} prematurely), as it
will probably happen in the intermediate steps of our proof. Hence, it is required in general
to define a new equality predicate and define it by means of equations. Finally, structured
specifications are possible, supporting \emph{protecting} (no ``junk'' and no 
``confusion'' are allowed, as explained in the previous section), \emph{extending}
(no ``confusion'' is allowed), and \emph{using} (both ``junk'' and ``confusion'' are allowed).

Proof scores are defined in open-close environments. These environments
allow users to import and enrich other modules and execute reduction commands
without creating a new module. For example, a single proof score is usually enough
for discharging the base case because no case splittings 
(and hence new equations) are required, but the rest of constructors require
several open-close environment to distinguish different cases.

\subsection{CafeOBJ}\label{subsec:lisp_imp}

CafeOBJ~\cite{cafeReference2018,cafe-report} is a programming language
that inherits OBJ3 features and philosophy and adds support for hidden 
algebras~\cite{hiddenAlgebra} and rewriting logic~\cite{Marti-OlietMeseguer02b},
which can be combined with order-sorted algebra. This combined approach
has a formal mathematical foundation based in multiple 
institutions~\cite{cafe-report}. Besides these new theoretical features, 
CafeOBJ is implemented in Lisp and provides a better performance than
OBJ3, new predicates, such as a \texttt{search} predicate to test reachability,
and new tools, such as an inductive theorem prover~\cite{cafeReference2018}.
Because CafeOBJ is currently the language with the best support for proof scores,
we will use it for all the examples in this paper. We present below some details of
the syntax, which has been introduced in the previous section; a complete example
will be discussed in Section~\ref{subsec:ots}.

CafeOBJ supports modules with tight semantics, with syntax \texttt{mod! MOD-NAME \{...\}},
and loose semantics, with syntax \texttt{mod* MOD-NAME \{...\}}. Both types of modules can
be parameterized and the same importation modes presented for OBJ3 can be used in CafeOBJ.
Sorts are enclosed in square
brackets, while the subsort relation is stated by means of \verb"<".
Operators use the keyword \verb"op" (\verb"ops" if several operators are defined at the same time),
followed by the operator name, \verb":", the arity, \verb"->", the coarity, and a set of attributes
enclosed in curly braces; among these attributes we have \verb"ctor" for constructors, \texttt{assoc}
for associativity, \texttt{comm} for commutativity, and \verb"id:" for identity. The behavior of functions
is defined by means of equations, with syntax \verb"eq" (\texttt{ceq} for conditional equations).
It is worth remarking that
CafeOBJ provides a predefined \verb"_=_" predicate for all sorts and a
default equation indicating that it holds for terms syntactically equal. The user can enrich this
definition by adding extra equations.

Proof scores, with syntax \texttt{open MOD-NAME . STMS close}, take the definitions in the module \texttt{MOD-NAME}
and extend them with the statements in \texttt{STMS}. These statements include fresh constants, premises
(possibly using the \texttt{:nonexec} attribute to prevent them from being used in reductions), and reduce
commands, with syntax \texttt{red}.

\subsection{Maude}\label{subsec:maude}

Maude~\cite{maude-book} is a programming language implementing rewriting 
logic~\cite{Marti-OlietMeseguer02b}, a logic of change that is parameterized by an equational
logic. In Maude this logic is membership equational logic~\cite{BouhoulaJouannaudMeseguer00},
an extension of order-sorted equational logic that allows specifiers to state terms as having a given sort.
It supports parameterization, structured specifications, equational axioms, and reduction via rewriting.
Maude has been widely used for specifying and verifying systems, mainly focusing on its built-in
LTL model checker~\cite{maude-book}.

However, Maude does not support proof scores explicitly.
When no extra constants or equations are needed, for example for 
base cases, we can just select the appropriate module and reduce
the term.
Otherwise, we need to create new modules, import the specification
(Maude only allows theories to be imported in \emph{including} mode,
which allows junk and confusion to be added) and add constants and
equations in the usual way. Then, reduce commands can be used outside
the modules.
This way of defining proof scores, although very similar to the ones we have
already discussed, is more verbose. Because a major advantage of proof scores
is their flexibility and its ease of use, this extra effort might make Maude
inconvenient for large proofs.
Moreover, as OBJ3, Maude does not include a predefined equality predicate and
it must be explicitly defined. 
A detailed comparison between CafeOBJ and Maude can be found in~\cite{cafeInMaudeFAC}.

\subsection{CafeInMaude}\label{subsec:cafeinmaude}

CafeInMaude~\cite{cafeInMaudeFAC} is a CafeOBJ interpreter implemented in 
Maude~\cite{maude-book}. It supports non-behavioral CafeOBJ specifications,
open-close environments (and hence proof scores), and theorem proving.
Because CafeInMaude supports CafeOBJ modules and open-close environments,
it is possible to use exactly the same code presented in 
Section~\ref{subsec:lisp_imp}. However, CafeInMaude provides extra features,
discussed in Section~\ref{subsec:cimpx}, when proof scores are identified
with the \texttt{:id(...)} annotation.

\subsection{Comparison between implementations}\label{subsec:comparison}

Table~\ref{tab:comparison} summarizes the main aspects of the implementations above, where we
have excluded OBJ3 because it is discontinued. We have considered, for each implementation, the
following features:
\begin{itemize}
\item
The implementation language, which in some cases might limit the performance.

\item
The support for open-close environments in the language.

\item
The support for hidden algebras~\cite{hiddenAlgebra}, which are algebraic structures 
defined through their observable properties, focusing on external behavior without 
exposing their internal implementation.

\item
The existence of a search predicate, which has been used for exploring the
state-space in specific kinds of proof scores~\cite{DBLP:conf/birthday/Futatsugi15}.

\item
The existence of a search command, which has been used for checking invariant properties.

\item
The possibility of using model checking~\cite{clarke-model-checking-book} for analyzing linear temporal logic (LTL) properties on the specifications.
In this way, safety properties can be easily analyzed.

\item
The support of narrowing~\cite{narrowingFay,MeseguerThati04}, a generalization of term rewriting that allows logical variables in terms, replacing pattern matching by unification. 
This allows systems for symbolically evaluating these terms.

\item
The existence of a meta-level, where modules can be used as standard data, hence allowing
developers to reason about them.
\end{itemize}

Although some of these features are not directly related to proof scores, they also illustrate the
potential of these tools to improve their current approaches and their power as integrated
ecosystems for specifying and analyzing the properties of systems.

CafeOBJ is the only language implementing hidden algebra, which makes this implementation
the most appropriate for designers following this methodology; on the other hand, its Lisp implementation
limits its performance and makes it difficult to extend.
We also realize that Maude is
the most developed system, where novel approaches for verification have been implemented
lately (as discussed in Section~\ref{sec:open}, narrowing can be used to partially automate
proofs). However, the lack of open-close environments prevents users from using proof scores
at their full capability. CafeInMaude stands as an intermediate option: it supports open-close environments
and the search predicate, making it appropriate for using proof scores. Moreover, it is implemented in
Maude, which eases the integration of Maude features, such as narrowing, into proof scores.

\begin{table}[t]
\caption{Comparison between implementations}\label{tab:comparison}
\begin{center}
\begin{tabular}{ |c|c|c|c|c|c|c|c|c|c| } 
 \hline
  & Imp.\  & Open-close & Hidden & Search & Search & Model & Narrowing & Metalevel \\ 
  & lang.\ & environment   & algebra & predicate & command & checking & & \\
 \hline
 CafeOBJ         & Lisp     & Yes & Yes & Yes & No & No & No & No \\ 
 \hline
Maude             & C++     & No  & No & No & Yes & Yes & Yes & Yes \\ 
 \hline
 CafeInMaude & Maude & Yes & No  & Yes & No & Yes & No & Yes \\ 
 \hline
\end{tabular}
\end{center}
\end{table}

\section{Proof scores in practice}\label{sec:success}

In this section, we first present the main approach to proof scores: verification of observational
transition systems. We then discuss how new implementations and tools for CafeOBJ have
improved the analysis of this kind of systems.

\subsection{Observational Transition Systems}\label{subsec:ots}

Observational transition systems (OTS)~\cite{OgataF03FMOODS,OgataF06Goguen} are a general
methodology for formally specifying distributed systems. The intuitive idea is to model
system interactions by means of \emph{transitions}, while the values of the relevant 
components of the system are obtained by means of \emph{observations}.

The general notions of OTSs were first developed for behavioral 
specifications~\cite{logicalFoundationsCafe}, which distinguished two different types of sorts:
\emph{visible sorts}, used to define the data structures, and \emph{hidden sorts},
used to define the system. Because in these specifications the system is hidden it 
can only be analyzed via observations that show the value of particular components at 
each particular moment.
This approach was based on \emph{hidden algebras}~\cite{hiddenAlgebra}, which thanks to its
co-algebraic nature allowed specifiers to prove both invariant and liveness properties.
The constructor-based approach (see Section~\ref{sec:theory} for details) followed here
is limited to invariant properties.

The elements required for defining an OTS are:
\begin{itemize}
\item
A sort \texttt{Sys} standing for the system.

\item
A set of sorts $\mathcal{S}$ standing for the auxiliary data structures used by
\texttt{Sys}.

\item
The transitions available for the system. These transitions are defined as constructor
functions that return terms of sort \verb"Sys". Following these ideas, a standard 
definition of \texttt{Sys} with an initial state and $n$ transitions looks like:

\begin{itemize}
\item
\texttt{op init : -> Sys \{constr\}}

\item
\texttt{op} $\mathtt{trans_1}$ \texttt{: Sys} $\mathtt{Sort^1_1\ \cdots\ Sort^1_{k_1}}$
\texttt{-> Sys \{constr\}}

\item
$\cdots$

\item
\texttt{op} $\mathtt{trans_n}$ \texttt{: Sys} $\mathtt{Sort^n_1\ \cdots\ Sort^n_{k_n}}$
\texttt{-> Sys \{constr\}}
\end{itemize}

With all sorts $\mathtt{Sort^i_j} \in \mathcal{S}$, with $i \in \{1, \ldots, n\}, 
j \in \{1, k_i\}$. 
In general, for each transition (except for \texttt{init}) to
be triggered a condition $c_{\mathit{trans_i}}, i \in \{1, \ldots, n\}$, must be met. 
This condition is the so-called \emph{effective condition}; when it is not met, 
we have
$\mathtt{trans_i}(s, x_1, \ldots, x_{k_i}) = s$, for \texttt{s} a variable of sort
\texttt{Sys} and $x_1, \ldots, x_{k_i}$ variables of sort $\mathtt{Sort^i_1\ \cdots\ Sort^i_{k_i}}$,
respectively.

\item
Observation functions, which return the value of the data structures in the different
states. These functions are of the form:
\begin{itemize}
\item
\texttt{op} $\mathtt{o_1}$ \texttt{: Sys} $\mathtt{Sort^1_1\ \cdots\ Sort^1_{l_1}}$
\texttt{->} $\mathtt{Sort_1}$

\item
$\cdots$

\item
\texttt{op} $\mathtt{o_m}$ \texttt{: Sys} $\mathtt{Sort^m_1\ \cdots\ Sort^m_{l_m}}$
\texttt{->} $\mathtt{Sort_m}$
\end{itemize}

With all sorts $\mathtt{Sort^i_j} \in \mathcal{S}$,
$\mathtt{Sort_i} \in \mathcal{S}$, with $i \in \{1, \ldots, m\}, 
j \in \{1, l_i\}$, and $\mathtt{Sort_i}$ different to \texttt{Sys}. 

\item
For each observation function $o_i$ ($i \in \{1, \ldots, m\}$)
we define equations for each transition, assuming the effective condition holds:
\begin{itemize}
\item
\texttt{eq} $\mathtt{o_i(init, x^i_1, \cdots, x^i_{l_i}}) = v$

\item
\texttt{ceq} $\mathtt{o_i(trans_1(s, y^1_1, \cdots, y^1_{k_1}), x^i_1, \cdots\ x^i_{l_i}}) =
f(s, y^1_1, \cdots, y^1_{k_1}, x^i_1, \cdots\ x^i_{l_i})
\textrm{ if } c_{\mathit{trans_1}}$

\item
$\cdots$

\item
\texttt{ceq} $\mathtt{o_i(trans_n(s, y^n_1, \cdots, y^n_{k_n}), x^i_1, \cdots\ x^i_{l_i}}) =
f(s, y^n_1, \cdots, y^n_{k_n}, x^i_1, \cdots\ x^i_{l_i})
\textrm{ if } c_{\mathit{trans_n}}$
\end{itemize}

With $x^i_j$ ($j \in \{1, \ldots, l_i\}$) and $y^p_q$ ($p \in \{1, \ldots, n\}, 
q \in \{1, \ldots, k_p\}$) variables of sort $\mathtt{Sort^i_j}$ and 
$\mathtt{Sort^p_q}$, respectively, $v$ a constant of sort $\mathtt{Sort^i}$, and
$f$ an auxiliary function (possibly an observation function or a data structure)
not involving any transition.
\end{itemize}

In the following we detail how to define OTSs and how to verify them using proof
scores by means of the Qlock example. Qlock is a mutual exclusion protocol based
on a queue with atomic operations and a binary semaphore. The basic idea is that
processes can be in \emph{remainder section}, \emph{waiting section}, and
\emph{critical section}. When a process in the remainder section moves to the
waiting section its process identifier is added to the queue; once it reaches
the top of the queue it moves into the critical section. Processes exiting the critical section return to the
remainder section and the top element is removed from the queue. Although this system is not very complex, proving mutual
exclusion requires a proof score of approximately 350 lines of code and illustrates
some interesting features, as we will see below.

To specify this system as an OTS we need to identify the transitions and the
values we want to observe. It is clear in this case that transitions take place
when the processes change the state, while we want to observe, for each process
identifier, the state the process is in.
Hence, we start our specification by defining the auxiliary data structures. The 
module \verb"LABEL" for labels was already presented in Example~\ref{ex:label}.
The module \texttt{PID} defines the process identifiers: we define the sort \texttt{Pid}
for process identifiers, \texttt{ErrPid} for erroneous process identifiers, and
\verb"Pid&Err" as a supersort including both correct and erroneous identifiers.
This module also indicates that correct and erroneous process identifiers are
different:

{\codesize
\begin{verbatim}
mod* PID {    [ErrPid Pid < Pid&Err]
              op none : -> ErrPid
              var I : Pid
              var EI : ErrPid
              eq (I = EI) = false . }
\end{verbatim}
}

We define next a module \texttt{TRIVerr} with loose semantics to define the elements 
of a generic
queue, which will be later instantiated with the appropriate sorts. Because some
functions on queues are partial, we define the existence of both correct elements
(\texttt{Elt}) and erroneous elements (\texttt{ErrElt}), and a supersort enclosing
them. The module also defines a particular element of sort \texttt{ErrElt}, 
\texttt{err}.

{\codesize
\begin{verbatim}
mod* TRIVerr {   [ErrElt Elt < Elt&Err]
                 op err : -> ErrElt}
\end{verbatim}
}

The module \texttt{QUEUE} below is parameterized by \texttt{TRIVerr}. It defines
the sort for the empty queue (\texttt{EQueue}), non-empty queues (\texttt{NeQueue}),
and general queues (\texttt{Queue}) with the appropriate subsort relation between
them. The constructors for queues are \texttt{empty}, which builds the empty queue,
and \verb"_|_", which given an element and a general queue returns a non-empty
queue. The functions for enqueueing (\texttt{enq}) and dequeueing (\texttt{deq}) 
are total and defined as usual; for the particular case of dequeueing an empty 
queue the function returns the same empty queue, as shown on the right. The function 
for obtaining the \texttt{top} element of a queue is also total when using the error
sorts defined above: the top element for an empty queue is \texttt{err}:

{\codesize
\begin{verbatim}
 mod! QUEUE(E :: TRIVerr) {
   [EQueue NeQueue < Queue]                           var Q : Queue
   op empty :             -> EQueue {constr}          vars X Y : Elt.E
   op _|_   : Elt.E Queue -> NeQueue {constr}         eq enq(empty,X) = X | empty .
   op enq : Queue Elt.E -> NeQueue                    eq enq(Y | Q,X) = Y | enq(Q,X) .
   op deq : Queue -> Queue                            eq deq(empty) = empty .
   op top : EQueue -> ErrElt.E                        eq deq(X | Q) = Q .
   op top : NeQueue -> Elt.E                          eq top(empty) = err .
   op top : Queue -> Elt&Err.E                        eq top(X | Q) = X .   }
\end{verbatim}
}

We relate, by means of the view \texttt{TRIVerr2PID}, the elements in \texttt{TRIVerr}
and the ones in \texttt{PID} in the expected way:

{\codesize
\begin{verbatim}
view TRIVerr2PID from TRIVerr to PID {
  sort Elt -> Pid,
  sort ErrElt -> ErrPid,
  sort Elt&Err -> Pid&Err,
  op err -> none   }
\end{verbatim}
}

We use this view to instantiate the queue in the \verb"QLOCK" module, which also
includes the modules \texttt{LABEL} and \texttt{PID}:

{\codesize
\begin{verbatim}
mod* QLOCK {
  pr(LABEL + PID)
  pr(QUEUE(E <= TRIVerr2PID))
\end{verbatim}
}

We define the sort \texttt{Sys}, standing for the system, and the corresponding
transitions building it. In addition to \texttt{init}, which stands for the initial
state, we define the transitions \texttt{want} (for a process moving from the
remainder section to the waiting section), \texttt{try} (for processes in waiting
section trying to enter the critical section), and \texttt{exit} (for processes
in the critical section leaving it):

{\codesize
\begin{verbatim}
  [Sys]
  op init : -> Sys {constr}
  op want : Sys Pid -> Sys {constr}
  op try  : Sys Pid -> Sys {constr}
  op exit : Sys Pid -> Sys {constr}
\end{verbatim}
}

In our system we are interested in the label corresponding to each process
and the queue, so we define the appropriate observations to retrieve its value
in a given system:

{\codesize
\begin{verbatim}
  op pc    : Sys Pid -> Label
  op queue : Sys -> Queue
\end{verbatim}
}

We define some variables and define the observations for each state. In the
case of the initial state all processes are in the reminder section and the
queue is empty:

{\codesize
\begin{verbatim}
  var S : Sys      vars I J : Pid
  eq pc(init,I)  = rs .
  eq queue(init) = empty .
\end{verbatim}
}

For the transition \texttt{want} we define the effective condition first. In this
case to enter into the waiting section is only required that the process is in the
remainder section:

{\codesize
\begin{verbatim}
  op c-want : Sys Pid -> Bool
  eq c-want(S,I) = (pc(S,I) = rs) .
\end{verbatim}
}

When the condition holds the label corresponding to the process is \texttt{ws},
while the queue is modified by introducing the process identifier. When the
condition does not hold the transition is not applied:

{\codesize
\begin{verbatim}
  ceq pc(want(S,I),J)  = (if I = J then ws else pc(S,J) fi) if c-want(S,I) .
  ceq queue(want(S,I)) = enq(queue(S),I)                    if c-want(S,I) .
  ceq want(S,I)        = S                                  if not c-want(S,I) .
\end{verbatim}
}

We would proceed in a similar way with the rest of transitions.
Finally, we define the properties we want to prove. The property \texttt{inv1}
stands for mutual exclusion, while \texttt{inv2} states that the process in the
critical section must be the one on the top of the queue:

{\codesize
\begin{verbatim}
  op inv1 : Sys Pid Pid -> Bool
  op inv2 : Sys Pid -> Bool
  eq inv1(S:Sys,I:Pid,J:Pid) = (((pc(S,I) = cs) and pc(S,J) = cs) implies I = J) .
  eq inv2(S:Sys,I:Pid) = (pc(S,I) = cs implies top(queue(S)) = I) . }
\end{verbatim}
}

In the simplest case, we would just create an open-close environment with both
properties and, if both are evaluated to \texttt{true}, then we would know they
hold:

{\codesize
\begin{verbatim}
open QLOCK .
  op  s : -> Sys .
  ops i j : -> Pid . 
  red inv1(S:Sys, I:Pid, J:Pid) .
  red inv2(S:Sys, I:Pid) .
close
\end{verbatim}
}

As expected, they are not reduced to \texttt{true} and we need to inspect
each possible transition (that is, applying induction) and each property 
separately. The proof for \texttt{init} is straightforward; it is enough 
to just reduce the term:

{\codesize
\begin{verbatim}
  open QLOCK .                          open QLOCK .
     ops i j : -> Pid .                    op i : -> Pid .
     red inv1(init,i,j) .                  red inv2(init,i) .
  close                                 close
\end{verbatim}
}

We focus next on how to prove \texttt{inv1} for the transition \texttt{want},
the rest of the proof follows the same ideas and is available in the
repository above. The first approach for proving this property would be the
open-close environment below:

{\codesize
\begin{verbatim}
open QLOCK .
   op s : -> Sys .
   ops i j k : -> Pid .
   eq [:nonexec] : inv1(s,I:Pid,J:Pid) = true .
   eq [:nonexec] : inv2(s,I:Pid) = true .
   red inv1(want(s,k),i,j) .
close
\end{verbatim}
}

Note that it contains the induction hypotheses as non-executable equations,
so they are only used for informative purposes. When executed, we obtain the
following result:

{\codesize
\begin{verbatim}
Result: true xor cs = pc(want(s,k),i) and cs = pc(want(s,k),j) xor i = j and 
        cs = pc(want(s,k),i) and cs = pc(want(s,k),j) : Bool
\end{verbatim}
}

We could try to use the inductive hypotheses to further reduce it, but it has
no effect.
Then, we notice that the terms \texttt{pc(want(s,k),i)} and \texttt{pc(want(s,k),j)}
are not being reduced. This happens because there is no information about the
effective condition; in order to solve it we can indicate it holds, adding the
corresponding equation to the environment:\footnote{Notice that other options are
also available, being the simplest one adding the equation \texttt{i = j}; the user
is in charge of exploring the different alternatives and choosing the most
appropriate ones.}

{\codesize
\begin{verbatim}
open QLOCK .
   op  s : -> Sys .
   ops i j k : -> Pid .
   eq [:nonexec] : inv1(s,I:Pid,J:Pid) = true .
   eq [:nonexec] : inv2(s,I:Pid) = true .
   eq pc(s,k) = rs .
   red inv1(want(s,k),i,j) .
close
Result: true xor cs = if i = k then ws else pc(s,i) fi and cs = if j = k then ws else pc(s,j) fi xor i = j and 
        cs = if i = k then ws else pc(s,i) fi and cs = if j = k then ws else pc(s,j) fi : Bool
\end{verbatim}
}

In this case we need to state an equality between
the fresh constants. For example, we can add the equation \texttt{i = k}:

{\codesize
\begin{verbatim}
open QLOCK .
   op  s : -> Sys .
   ops i j k : -> Pid .
   eq [:nonexec] : inv1(s,I:Pid,J:Pid) = true .
   eq [:nonexec] : inv2(s,I:Pid) = true .
   eq pc(s,k) = rs .
   eq i = k .
   red inv1(want(s,k),i,j) .
close
Result: true : Bool
\end{verbatim}
}

Although we have obtained \texttt{true}, the goal is not completely
proven because it depends of some equations and we do not know 
what happens if they do not hold. For example, we would
need to use \texttt{(i = k) = false} instead of \texttt{i = k}, which
is not reduced to \texttt{true} and would require a reasoning similar to
the one above.

This proof illustrates the strong points of proof scores: it is possible to
prove the subgoals using the same syntax employed to specify the system. Although in this
example the equations that we needed to add were simple, it is also possible to define
more complex ones, possibly involving equational attributes such as associativity.
Moreover, the result obtained when reducing the environment guides the proof, so the
user must employ his/her expertise to choose the most promising ones.

On the other hand, this example also illustrates their weak point: the user is in charge 
of ensuring that all possible cases have been traversed; if one subgoal is not taken 
into account none of the tools will warn the user, which might be specially problematic 
with large proofs. This (partial) lack of formality is one of the reasons
why proof scores have not been more widely applied, while theorem proving, which has
a very strict syntax but also provides full confidence, is very popular. 
However, the equations added in open close-environments are not arbitrary: they are
case splittings and follow some rules. For this reason, we should be able to 
find a general schema for a large class a proof scores, so we can automatically generate
a proof in a ``standard'' theorem prover. In the next section we present a proof assistant
providing several features implicitly used in proof scores and a proof generator that,
given a proof score, tries to replicate the proof using the proof assistant.

\subsection{The CafeInMaude Proof Assistant and Proof Generator}\label{subsec:cimpx}

The CafeInMaude implementation of CafeOBJ
eases the specification of systems and the definition
of proof scores; in particular, it takes advantage of the underlying Maude 
implementation and its meta-level features to:
\begin{enumerate}
\item
Support extra features in CafeOBJ specifications. In particular, the current
version of CafeInMaude supports the \texttt{owise} (standing for \emph{otherwise})
attribute for equations, which indicates that the corresponding equation is only
used if the rest of equations cannot be executed.

\item
Extend the syntax in an easy way. CafeInMaude is implemented in Maude, taking
advantage of its powerful metalevel. In this way, the CafeOBJ grammar is defined
as a Maude module used for parsing, while the translation into Maude is performed
using equations.
Because Maude's and CafeOBJ's syntax and semantics are very similar, CafeOBJ
programmers can easily add new elements to the system to experiment with them.
Once these elements have been thoroughly tested they might become part of the
standard CafeInMaude distribution.

\item
Connect with other Maude tools. CafeInMaude translates CafeOBJ specifications into
Maude specifications, so any tool dealing with Maude specifications can be 
connected with CafeInMaude. In particular, it is interesting to consider the
Maude Formal Environment~\cite{mfe07}, which includes a termination checker, 
a confluence checker, a coherence checker, and a theorem prover, and the Maude
integration with SMT solvers.

\item
Execute larger proofs than the Lisp implementation of CafeOBJ.
As shown in~\cite{cafeInMaudeFAC}, the performance of CafeInMaude is much better
than the one provided by the CafeOBJ implementation described in
Section~\ref{subsec:lisp_imp},
especially when the \texttt{search} predicate is used.
This allows CafeInMaude to prove properties on some protocols that fail in CafeOBJ
because the interpreter runs out of memory.
\end{enumerate}

\subsubsection{The CafeInMaude Proof Assistant}\label{subsubsec:cimpa}

Beyond the features above, CafeInMaude provides a theorem prover, the CafeInMaude
Proof Assistant (CiMPA)~\cite{tosem18}, supporting:
\begin{itemize}
\item
Induction in one or more variables.

\item
Instantiating free variables with fresh constants (i.e., applying the theorem of constants).
In this way, we can reason with the most appropriate values of the variables for the
particular subgoal.

\item
Splitting the goal. When the current goal is composed of several equations, it
is useful to discharge each of them separately.

\item
The use of the induction hypotheses as premises. We can use an implication with one
of the induction hypotheses as premise to simplify the current subgoal.

\item
Case splitting by (i) true/false, (ii) constructors, and (iii) special combinations
for associative sequences.

\item
Discharging the current subgoal by using reduction.
\end{itemize}

\subsubsection{The CafeInMaude Proof Generator}\label{subsubsec:cimpg}

While in CafeOBJ proof scores and proof
scripts are unrelated entities (from the tool support point of view), CafeInMaude 
relates them by means of the CafeInMaudeProofGenerator (CiMPG), which makes
proof scores a \emph{formal} (rather than semi-formal) proof technique. 

Each subgoal in a CiMPA proof extends
the original CafeOBJ module with the assumptions (and the corresponding fresh
constants needed to define them) generated due to induction and case splitting.
In the same way, each open-close environment consists of a module extended with
fresh constants and equations. Hence, it should be possible to relate, up to
substitution, both of them and discharge the corresponding subgoal using
reduction, possibly using first an implication with a substitution extracted
from the open-close environment.

In addition to using induction at the beginning and, possibly, the theorem
of constants, the CiMPG algorithm needs to build the sequence of case splittings. 
In order to generate this sequence it must compute (i) the kind of case splitting
applied and (ii) the order in which the case splittings are applied. The details
of the algorithm are complex and are described in detail in~\cite{tosem18}.

If CiMPG is successful, 
a formal proof for the theorems is obtained; otherwise, it points out those subgoals
whose proof could not be generated, which helps users to find errors in proof scores.
In this way the user can keep using proof scores, taking advantage of its flexible
nature, while leaving the verification of the proof to CiMPG.

In~\cite{DBLP:journals/jss/RiescoO22} the CafeInMaude Proof Generator and Fixer-Upper (CiMPG+F),
a tool for generating complete CiMPA proof scripts from incomplete proof scores, is 
presented. CiMPG+F follows a bounded depth-first strategy for applying cases splitting
to subgoals until they are discharged or the bound is reached. The candidates for
case splitting are, roughly speaking, chosen by reducing at the metalevel the subgoals
and picking those terms that were not be reduced even though they were not built
by using constructor functions.

\subsection{Invariant Proof Score Generator}\label{subsec:ipsg}

Using CafeInMaude, another tool called Invariant Proof Score Generator (IPSG)~\cite{tls12} has been implemented, which can automate the proof score writing process. Precisely, given a CafeOBJ formal specification, invariant properties formalizing the properties of interest together with an auxiliary lemma collection, IPSG can generate the proof scores verifying those properties.

Returning to the Qlock example introduced in Section~\ref{subsec:ots}, to prove the two properties \texttt{inv1} and \texttt{inv2}, we use the following script:

{\codesize
\begin{verbatim}
ipsgopen QLOCK .
  inv inv1(S:Sys, I:Pid, J:Pid) .
  inv inv2(S:Sys, I:Pid) .
  generate inv1(S:Sys, I:Pid, J:Pid) induction on S:Sys .
  generate inv2(S:Sys, I:Pid) induction on S:Sys .
close
set-output proof-scores.cafe .
:save-proof . 
\end{verbatim}
}

\noindent
Feeding this script into IPSG, the tool will generate the proof scores and save them to file ``\texttt{proof-scores.cafe}''.

How proof scores are generated can be briefly summarized as follows. 
Starting from a collection of open-close fragments, where each of them does not contain any equation and the most typical induction hypothesis instance is used if it is an induction case,
IPSG uses Maude metalevel functionalities to reduce the goal to its normal form.
There are three possible cases:

\begin{itemize}
	\item The obtained result is \texttt{true}. The proof is simply discharged.
	
	\item The obtained result is \texttt{false}. IPSG then tries to discharge the associated case by finding a suitable lemma instance based on the lemma collection provided by human users. The key idea is to enumerate all lemma instances constructed from the terms existing in the current open-close fragment to check if it can discharge the proof. 
	
	\item The obtained result is $x$, which is neither \texttt{true} nor \texttt{false}.
	A sub-term of $x$, let's say $x'$, will be then chosen by IPSG to split the current case associated with that open-close fragment into two sub-cases: one when $x'$ holds and the other when it does not. 
    The same procedure is applied for each sub-case produced until either \texttt{true} or \texttt{false} is returned for the reduction. 
\end{itemize}

Which sub-term $x'$ is chosen at each step determines the order of case splittings, which is a significant factor affecting the tool's efficiency. 
Some heuristics techniques were implemented for choosing case splitting order.
For each induction case, the highest priority is given to those case splittings to reduce the effective condition of the associated transition to either true or false. 
If term $x$ contains a sub-term ${\bf if} \ c \ {\bf then} \ a \ {\bf else} \ b \ {\bf endif}$, the sub-terms of the condition $c$ are given higher priority than others.
In summary, IPSG can automatically conduct case splitting such that true or false is returned for each sub-case. 
Human users, therefore, only need to concentrate on the most difficult task in interactive theorem proving, i.e., to conjecture lemmas. 

IPSG does not support case splitting based on sort constructors as CiMPA does.
To confirm the soundness of the proof scores produced by IPSG, CiMPG is first used to generate proof scripts from the proof scores, and the proof scripts are executed with CiMPA.
The finding lemma process may take time if a single lemma instance cannot discharge a case but more than one lemma instance are needed.

\textbf{Discussion.}
Among the challenges that proof scores face, two of them are due to their reliance in
human involvement: missing open-close environments and making use of strategies
unsupported by the theory (e.g.\ using \texttt{false} as a premise, which automatically
discharges a goal). Using a theorem prover such as CiMPA (Section~\ref{subsubsec:cimpa})
solves these problems but takes us out of the proof scores approach. CiMPG 
(Section~\ref{subsubsec:cimpg}) relates
proof scores and theorem provers (CiMPA in this case), so proof scores can be validated.

CiMPG+F (Section~\ref{subsubsec:cimpg}) and IPSG (Section~\ref{subsec:ipsg}) do
not face these challenges, which are specific to proof scores, but more general challenges:
those faced by human users when producing a proof. In particular, they help finding case splittings
and when to use lemmas for discharging the current goal.

\section{Success stories}\label{sec:success_cases}

In this section we present several cases where proof scores have been
used to verify protocols and systems. First, we give a general view of
the different protocols/systems tackled.  Then, we present a
particularly interesting example that illustrates the power of the
approach.
We have grouped them by topic, so specifiers can find references
for the systems they are verifying.
In particular, proof scores have been used to specify and verify:

\begin{itemize}
\item Shared-memory mutual exclusion protocols -- Qlock (an abstract
  version of Dijkstra binary semaphore), Lamport bakery
  protocol\,\cite{OgataF08ICFEM}, MCS
  protocol\,\cite{mcs_access}, etc.

\item Distributed mutual exclusion protocols -- Suzuki-Kasami
  protocol\,\cite{OgataF02FMOODS} and Ricart-Agrawala
  protocol\,\cite{OgataF01APAQS}.

\item Communication protocols -- Alternating Bit Protocol
  (ABP)\,\cite{OgataF13JUCS} and Simple Communication
  Protocol (SCP).

\item Authentication protocols -- Needham-Schroeder-Lowe public-key
  (NSLPK) authentication
  protocol\,\cite{DBLP:journals/entcs/OgataF02,OgataF03FMOODS},
  Transaction Layer Security (TLS)\,\cite{DBLP:conf/icdcs/OgataF05},
  Timed Efficient Stream Loss Tolerant Authentication (TESLA)
  protocol~\cite{OuranosOS14IEICE}, etc.

\item Electronic commerce (e-commerce) protocols -- $i$KP electronic
  payment
  protocol\,\cite{DBLP:conf/isss2/OgataF02,Ogata2010IJSEKE},
  Mondex electronic purse protocol\,\cite{DBLP:conf/ifm/KongOF07},
  Secure Electronic Transactions
  (SET)\,\cite{Ogata2004QSIC}, etc.

\item Real-time systems: Fischer mutual exclusion
  protocol~\cite{OgataF07SCP,DBLP:conf/sice/0001HSO20},
  TESLA protocol~\cite{OuranosOS14IEICE}, etc.

\item Modern cryptographic protocols and post-quantum cryptographic protocols: 
TLS 1.2~\cite{tls12}, 
Hybrid Post-Quantum TLS and Hybrid Post-Quantum Secure Shell (SSH)~\cite{duong-thesis,10375488}.

\item Other protocols/systems: Electronic government (e-government)
  systems\,\cite{KongOF10IEICE},
  Workflows\,\cite{KongOF07IJSEKE}, Fault Tree
  Analysis\,\cite{DBLP:conf/IEEEcit/XiangFH04,DBLP:conf/qsic/XiangO05},
Internal control\,\cite{ArimotoIF12IEICE},
Dynamic update systems\,\cite{ZOF14SAS},
Context-aware adaptive systems\,\cite{KsystraSF15IJSEKE},
Digital right management systems\,\cite{TriantafyllouSF13IEICE},
Mobile systems\,\cite{DBLP:journals/ieicet/OuranosSF07},
Social network systems\,\cite{DBLP:journals/corr/abs-1106-6267},
Declarative cloud orchestration\,\cite{DBLP:conf/icfem/YoshidaOF15},
Hybrid (cyber-physical)
  systems\,\cite{DBLP:conf/ifip10-3/OgataYSF04,DBLP:journals/corr/abs-2010-15280}, 
  concurrent programs with safe memory reclamation concern\,\cite{DBLP:conf/wrla/TranO24}, etc.
\end{itemize}

In addition to full formal verification of systems properties with
proof scores, partial formal verification of such properties with proof
scores can be conducted, checking/testing the cornerstones of such
systems. Mutual exclusion has been fully verified for the Suzuki-Kasami 
protocol~\cite{mta24} and partially verified for the 
Ricart-Agrawala protocol~\cite{OgataF01APAQS}. It
is a piece of our future work to fully formally verify the latter
protocol; no serious challenges are foreseen for the complete proof. 
CiMPG has been successfully used for verifying 
Qlock, ABP, SCP, and NSLPK~\cite{tosem18}, TLS 1.0~\cite{tls10},
and TLS 1.2~\cite{tls12}. 
In addition to
invariant properties, leads-to properties, a class of liveness
properties, can be handled by proof
scores\,\cite{DBLP:journals/ieicet/OgataF08,DBLP:conf/lopstr/PreiningOF14},
but few studies on theoretical aspects of this direction have been
conducted, which is another piece of our future work.

\section{Open issues}\label{sec:open}

In this section we discuss some lines of future work that researchers 
are encouraged to follow in order to widen the range of applications that 
can be managed via proof scores, while making them suitable for a larger 
audience.
We start discussing why proof scores have not reached a large audience and,
based in these reflexions, explore some possible solutions.

\subsection{Discussion: Why proof scores have not been widely adopted}
\label{subsec:why}

In order to understand why proof scores are not widely used it is necessary
to understand the big picture, that is, why \emph{theorem proving}, as a
general philosophy, has not reached industry except for particularly
complex or critical problems. We will list in the following some general
problems and discuss how they affect proof scores.

Of course, theorem proving has several advantages~\cite{sommerville,guideFM}
besides proving the soundness of a system: it helps designers to understand 
the details of the system, it is transparent to changes in the implementation
details (in particular, to optimizations), and guides programmers to find 
the errors when a proof fails. In fact, some of the problems listed below
might be considered ``myths'' against theorem proving (and formal
methods in general), but they are shared by many people and must be 
answered effectively to reach the public.

\subsubsection{Theorem proving is difficult}
\label{subsubsec:difficult}

The main argument used against formal methods is that they are difficult.
There are many arguments supporting this idea:
\begin{itemize}
\item
In general, theorem proving requires a \emph{specification} of
the system being analyzed.
Specifications are written using a specification language, which is
in most cases different from the implementation language. Moreover,
it requires programmers to reason on a different abstraction level.

\item
Proofs in some cases have their own language (possible a large set
of commands) that is different from the specification and the implementation
languages.

\item
Once specifiers know the specification and the proof language, theorems
must be stated. This requires previous knowledge (mainly logic) and a deep
understanding of how the system behaves.

\item
Finally, the proof must be done. Proving a theorem requires a careful
planning, guessing the necessary lemmas, and choosing the most adequate
case splittings in each step.
\end{itemize}

In the particular case of proof scores, the specification language and
the proof language are the same, and it is flexible enough to ease the
demonstration process in some extent. Of course, finding the theorem
and the lemmas and proving them are intrinsic problems of theorem proving
and cannot be completely solved by any tool, so only partial solutions can
be found.

\subsubsection{There exists a gap between specifications and implementations}
\label{subsubsec:gap}

One of the main advantages of specifications is their ability to abstract the implementation 
details, focusing instead on their properties. However, this advantage can be also
understood as a weakness when trying to formally relate specifications and their
corresponding implementations. This weakness is specially important from the industrial
point of view, where the systems that require formal verification are those being really deployed.

In this direction, the $\mathbb{K}$-framework~\cite{k-framework} is the touchstone we should
follow. $\mathbb{K}$ is a framework where programming languages have formal definitions
and tools for a given language can be derived them. This framework was originally developed
in Maude, so it should be possible to build a similar architecture where, given the semantics
of a language and a proof, it can be ``lifted'' to the real system in a safe way.

\subsubsection{Theorem proving requires experts}
\label{subsubsec:experts}

As a consequence of the problems in Section~\ref{subsubsec:difficult},
it might seem that formal methods cannot be applied by standard engineers
but only by formal methods experts, specifically trained to deal with
this kind of tools. Hence, it might be worth hiring ``theorem proving 
professionals'' rather than training them. This problem is worsened
by the interfaces offered by theorem provers, because most of them do not
provide a graphical or web interface, while commercial languages are 
integrated into familiar development environments, which eases its use.

This problem might have an impact in proof scores. The specification
languages used for proof scores have solid mathematical foundations,
and proof scores themselves have a formal basis, as shown in
Sections~\ref{sec:theory} and~\ref{sec:spec},
and then potential users might consider only those that fully understand
the theory underlying them can use them. The truth is that, although
understanding some basic principles is needed to write correct proof
scores, this knowledge is much simpler than understanding the complete
framework, making the approach available in practice to a wide audience.
It is also true that the command-based interface offered by proof scores
is less intuitive than state of the art development environments, so
it requires time to get used to it.

\subsubsection{Theorem proving consumes (too much) resources}
\label{subsubsec:time}

In general, real developments have a tight schedule and a tight
budget, so devoting developers and time to formal methods is often
considered a ``waste.'' Even though it has been shown 
(see e.g.~\cite{amazonFM}) that formally analyzing a system benefits
the software development cycle, it also requires an initial effort
that small/medium software development companies (think that they) cannot 
afford.

As discussed in Section~\ref{subsubsec:difficult}, proof scores use
the same language for specifying and proving and have a very flexible
syntax to ease the verification process. Moreover, CafeOBJ
is a high-level language for fast prototyping, so the resource
consumption is reduced with respect to other proposals.
However, it is still true that users need to get used to the environment
and to the proof philosophy, so it will require an initial investment,
and time and resources will inevitably be consumed later for each 
development.

\subsubsection{Theorem proving is only useful for critical applications}
\label{subsubsec:critical}

As a consequence of the previous difficulties, theorem proving has been
relegated to very specific scenarios. However, even for these scenarios
it is usually the case that other non-formal techniques like testing
are preferred, because they are better known and integrated into the
software development cycle.

In fact, proof scores (and theorem proving in general) have been 
traditionally applied to protocols that must verify some critical properties. 
This makes sense because a failure in these systems might have a high 
economic cost, but it is also the case that any error in commercial software
produces a monetary loss and reduces the brand value, so it is worth
verifying as much software as possible. 

However, this problem is combined with the ones above, because the
intrinsic difficulty of theorem proving and the resource consumption
might make verification too expensive. For these reasons, more powerful
theorem proving frameworks are required, so it can be applied beyond
critical applications.

\subsubsection{Theorem proving does not work with real problems}
\label{subsubsec:real}

In order to apply theorem proving in real cases some features
are required: (i) the expressivity of the specification language is
adequate to represent even subtle cases in an easy way; (ii) the
prover is powerful enough, that is, it provides an acceptable level
of automation and it supports a wide enough range of commands to deal
with multiple situations; and (iii) if possible, the tool has been
already used to tackle real problems or it has been designed in collaboration
with industry.

In the case of proof scores, CafeOBJ has been developed in close contact
with industry~\cite{cafeIndustry}. However, real maturity will be reached
if they keep being used once collaboration with academia finishes.

\subsubsection{Not even theorem proving provides complete assurance}
\label{subsubsec:complete}

A critical issue in theorem proving (and formal methods in general) is the
correspondence between the specification and the implementation.
Although some theorem provers, such as Coq\,\cite{DBLP:series/txtcs/BertotC04} and Agda\,\cite{DBLP:conf/tphol/BoveDN09}, support translating their formal proofs or specifications to executable programs written in other programming languages, such as Haskell, the generated programs are functional but neither are imperative nor concurrent.
There also exist studies on translating CafeOBJ specifications based on OTS to Java programs\,\cite{DBLP:conf/apsec/HaO17, DBLP:conf/seke/SenachakSOF05}, however, either the generated programs are sequential or humans need to write the programs manually based on the annotation inserted into the specifications.
Different conformance relations~\cite{tretmans93,tretmans96}
have been proposed, but it is not always the case that they can be easily applied.
Moreover, non-experts (see Issue~\ref{subsubsec:experts}) might incorrectly
define a property, hence making the proof meaningless.

In fact, a particular aspect of this problem affects proof scores because 
they are semi-formal in the sense that it is not checked in any way whether 
the proof score is sound, as discussed in the previous sections. This problem
has been mitigated by CiMPG, as described in Section~\ref{subsec:cimpx}.
However, the conformance problem remains open.

\subsubsection{Further discussion}

Regarding the issues above, it is interesting the description given by Amazon
in~\cite{amazonFM}, where they explain why they started using model checking 
in 2014 (and, in part, why they did not adopt it earlier) and how they are
benefitting from them. In fact, they highlight their necessity for a formal
tool that (i) is able to \emph{handle complex problems} (that is, able to
work with real problems, Issue~\ref{subsubsec:real}), minimizes cognitive
burden (that is, is not too difficult, Issue~\ref{subsubsec:difficult}, and 
does not require experts, Issue~\ref{subsubsec:experts}), and (iii) has
a high return in investment (that is, it does not consume too much resources,
Issue~\ref{subsubsec:time}). These restrictions are likely to be posed by
any other company and illustrate the problems (and, to some extent, the
\emph{myths}) of formal methods. In order to prove these prejudices wrong
it is required to implement tools that give an answer to them, as we will discuss
in the next section.

\subsection{Towards a new proof scores paradigm}

We discuss in this section possible improvements for proof scores and how
they (partially) solve the problems posed in the previous section.

\subsubsection{Automatizing lemma discovery and proofs}\label{subsec:auto}

There are two reasons that make proving theorems a difficult task:
(i) finding the most appropriate theorems/lemmas for each system and
(ii) proving these theorems/lemmas by finding the appropriate case
splittings and applying the induction hypotheses when required.
Although it is not possible to fully automate these areas, it is possible
in general
to help the user in different ways, ensuring in all cases that the user
is in charge of the creative tasks while the computer executes the
computation-intensive tasks.

Regarding the theorem and lemma discovery, a promising direction is using
a graphical, executable description of the problem to get a better understanding
of the system. This idea has been studied for mutual exclusion and
communication protocols in
the SMGA tool~\cite{mta24}
to deal in a better way with message generation and consumption.
SMGA uses Maude as underlying specification language and rewrite engine 
to define and execute the protocols; then, a particular trace of the execution
can be fed into SMGA to animate it.
A careful, sequential visualization of many different traces might help
the user to understand the system and suggest 
different properties, helping the user to detect lemmas.

The tool has a number of limitations: (i) it is not directly integrated
with Maude, but it requires a text file with the trace information to be
generated beforehand; (ii) it is not possible to visualize two traces at
the same time; and (iii) it is designed to display protocols with a particular
set of features, so it might not support protocols using particular data
structures. Improving the tool in these directions would greatly help users,
specially if it is integrated in an IDE, as discussed in the next section.

Regarding proofs, it should be possible to partially automatize the proof
of a given theorem. Although even for simple protocols the number of possible
case splittings and instantiations of the inductions hypotheses is huge,
it is possible to reduce it by using the current subgoal to direct a bounded
depth-first search and try to discharge particular subgoals. If not successful,
it would provide useful information of those paths that were tried but did
not lead to success.

A limited version of this idea was presented in~\cite{creme07}, where the
authors present Cr\`eme, a tool checking invariants in CafeOBJ OTS specifications 
by trying case splitting. However, it only worked for true-false case splittings,
hence making it useful only for a small set of examples.
Extended versions of the approach, integrated into CafeInMaude, have been investigated 
in~\cite{DBLP:journals/jss/RiescoO22} (see Section~\ref{subsubsec:cimpg}) 
and~\cite{tls12} (see Section~\ref{subsec:ipsg}).
The benchmarks show that these approaches are successful for 
medium-sized examples, but their potential might lie in the combination with other
tools (in particular, graphical interfaces), so it can be used on-demand for specific
parts of the proof.

The latest developments of unification and narrowing in Maude (see Section~\ref{subsec:comparison})
provide a new opportunity for automatization. The nuITP~\cite{nuItp} supports narrowing
simplification, which allows nuITP to simplify some terms, even if they are not ground,
by computing a unifier substitution that applies to both the current goal and the lefthand
side of the equation being applied. This technique has been used successfully for discharging
a large number of subgoals automatically.
This idea can be applied in proof scores in two different ways:
\begin{itemize}
\item
It can be used to avoid the instantiation of induction hypotheses in some cases: 
the reduction would be tried using unification instead of matching and then narrowing would be used.

\item
The idea would be similar for discovering case splittings or lemmas. Because we want
each open-close environment in a proof score to be reduced to true, when it is not the case,
tools like CiMPG+F~\cite{DBLP:journals/jss/RiescoO22} explore the equations, looking
for those that could not be applied, and propose case splittings based on them. Using
unification, we will find new, more general case splittings because grounds
terms will not be required.

Moreover, when the current goal is reduced to false and the known lemmas cannot be
used to discharge it, we would analyze the subterms in the goal to detect those producing
false and try to synthesize a new lemma, possibly based in the equations of the system.
If the given lemma can be reduced to false (using unification and narrowing to avoid unnecessary
instantiations) it would be proposed to the user.
\end{itemize}

\subsubsection{IDE and Graphical User Interface support}\label{subsec:ide}

Large scale formal tools, such as Isabelle/HOL~\cite{DBLP:books/sp/NipkowPW02} and 
Dafny~\cite{DBLP:conf/lpar/Leino10}, provide users with editors, debuggers, etc.\ or are even 
integrated with commercial Integrated Development Environments (IDE), like in 
the case of Dafny working under Visual Studio. The size of the toolset gives a
rough measure of the maturity of the tool and gives users confidence, so it is
worth to create the most appropriate ecosystem where proof scores can be
used at their best.

An attempt to integrate Maude into Eclipse is available in~\cite{mdt}. This 
integration, however, is very limited and only takes advantage of a limited
number of features: syntax is highlighted and the Maude console is displayed
to interact with it. However, it does not integrate the debugger and does not
auto-complete function/variable names, among other features available in standard
programming languages such as Java. 

A similar, more powerful integration would be very interesting for CafeOBJ.
It would ease the development of proof scores to those software engineers 
that are already used to this kind of IDEs. Moreover, it would also help expert
CafeOBJ programmers, allowing them to focus on the proofs while the IDE helps
with the low level details like the current proof tree and the current subgoal.
If combined with a tool for automatically generating proofs like the one
discussed above, it could show case-splitting candidates
and even faulty proof branches, which would prevent the user from using
case splittings leading to dead-ends.

Besides these general improvements, specific features to deal with proof scores,
proof scripts, and proof trees are required. In particular:
\begin{itemize}
\item
The proof tree should be depicted, indicating for each node the current subgoal,
the induction hypotheses, the fresh constants introduced thus far, and the case 
splittings used to reach the node.

\item
The relation between each open-close environment in a given proof scores and
the proof tree should be made explicit, if possible.

\item
Suggested case splittings and hypotheses instantiations, if a mechanism as the 
one proposed above is available.
It should also show which case splitting have been automatically tried and
discharged.

\item
Different traces leading to this subgoal; this might ease the lemma discovery
sketched in the previous section.

\item
The equations in the original specification involved in the goal, so the specification
can be revised if necessary.
\end{itemize}

\subsubsection{Application to New Protocols}\label{subsec:new_protocols}

During the last years the popularity of new paradigms based on big data, 
machine learning, neural networks, large language models,
the Internet of Things, and blockchain is gaining much attention from the
scientific community. The complexity of the algorithms being applied in 
these areas and the necessity of keeping the used data safe will require
a huge effort from the formal methods community, that must verify whether
different properties hold in these systems.

In order to specify and verify these systems it might be worth devising particular
techniques that abstract and automate certain components (as suggested in
Section~\ref{subsec:auto}), so it becomes easier to deal with some classes
of programs. For example, it would be worth defining a framework to analyze
the consistency of transactions that could be used for many different NoSQL databases.

\subsubsection{Artificial intelligence support}\label{subsec:ia}

The current developments in Artificial Intelligence (AI) and Large Language Models
make difficult to foresee how formal methods will evolve in the following years.
Nowadays, tools like GitHub Copilot~\cite{copilot} help programmers to develop
``standard'' software applications. Although the main impediment to implementing
this type of tools for specification and verification is the relatively low amount of 
data for training, recent advances in AI support specialization in particular fields
and training with ``small data,'' so in the following years formal methods support
will probably become universal. This support will include help for specifying the
system, finding the appropriate properties to verify, and discharging them.
Some works in this line include~\cite{refinementLLM}, a tool designed for Coq
for verifying program refinements; \cite{DBLP:journals/corr/abs-2311-07948},
a system for finding loop invariants; and the framework ALGO~\cite{ZhangWXWL23},
which allows programmers to synthesize and verify algorithms.

Once this kind of applications becomes a reality, some of the issues above
will be answered. In particular, the whole process will become easier
(Issue~\ref{subsubsec:difficult}); although experts will still be needed 
to supervise and interpret the results, Issue~\ref{subsubsec:experts}
will be solved to some extent; and the effort required (Issue~\ref{subsubsec:time})
in the whole process will be diminished.

\section{Related work}\label{sec:rel}

Several algebraic specification languages were developed from
around-80's through 2000's, among which are as follows: ACT
ONE\,\cite{EhrigFH83ADT},
CafeOBJ\,\cite{DiaconescuF98WS},
CASL\,\cite{DBLP:journals/tcs/AstesianoBKKMST02},
HISP\,\cite{DBLP:conf/ifip/FutatsugiO80},
Larch\,\cite{DBLP:series/mcs/GuttagHGJMW93},
Maude\,\cite{maude-book},
OBJ2\,\cite{DBLP:conf/popl/FutatsugiGJM85}, and
OBJ3\,\cite{Goguen99SEOBJ}. 
OBJ2, OBJ3, CafeOBJ, and Maude are members of the OBJ language
family. Static data can be specified in OBJ2 and OBJ3, while dynamic
systems can be specified as well in CafeOBJ and Maude.

Larch provides the Larch Prover
(LP)\,\cite{DBLP:conf/rta/GarlandG89}. Guttag et\ al.\ write the
following in the Larch book\,\cite{DBLP:series/mcs/GuttagHGJMW93}
about proving:

\begin{quote}
Proving is similar to programming: proofs are designed, coded,
debugged, and (sometimes) documented.
\end{quote}

\noindent
This ``proving as programming'' concept is shared by proving by proof
scores in OBJ languages. Although Larch and OBJ languages are based on
algebras, there are some differences. OBJ languages are pure-algebraic
specification languages, while Larch is not. OBJ languages adopt the
complete set of rewrite rules proposed by
Hsiang\,\cite{DBLP:journals/ai/Hsiang85} for propositional calculus,
while Larch does not. Larch uses the Knuth-Bendix completion
procedure, while OBJ languages do not. Note that OBJ languages use a
completion procedure to implement rewriting modulo associativity
and/or commutativity, adding some
equations~\cite{gog-tpa}.

There are several formal verification tools for Maude. Among them are
the Inductive Theorem Prover (ITP)\,\cite{DBLP:journals/jucs/ClavelPR06}
(including its latest version, the nuITP~\cite{nuItp}),
the Constructor-based Inductive Theorem Prover
(CITP)\,\cite{DBLP:conf/calco/GainaZCA13}, the Maude LTL model
checker\,\cite{DBLP:journals/entcs/EkerMS02}, the LTLRuLF model
checker\,\cite{DBLP:journals/scp/BaeM15}, and the Maude Invariant
Analyzer Tool (InvA)\,\cite{DBLP:conf/birthday/RochaM14}, where LTL
stands for linear temporal logic and LTLRuLF stands for linear
temporal logic of rewriting under localized fairness. Meseguer and
Bae\,\cite{DBLP:journals/scp/BaeM15} proposed a tandem of logics that
is the pair $(\mathcal{L}_S, \mathcal{L}_P)$ of logics, where
$\mathcal{L}_S$ is the logic of systems and $\mathcal{L}_P$ is the
logic of properties. The tandems of logics for these five tools are
$({\rm OSEL}, {\rm OSEL})$, $({\rm CbOSEL}, {\rm OSEL})$, $({\rm RWL},
{\rm LTL})$, $({\rm RWL}, {\rm LTLR})$, and $({\rm RWL}, {\rm OSEL})$,
respectively, where OSEL, CbOSEL, RWL, and LTLR stand for order-sorted
equational logic, constructor-based order-sorted equational logic,
rewriting logic, and linear temporal logic of rewriting,
respectively. As indicated by the names, the first two tools are
theorem provers, while the next two tools are model checkers. The
fifth one is a theorem prover.

CASL has a tool support called the Heterogeneous Tool Set
(Hets)\,\cite{DBLP:conf/cade/MossakowskiML07}. Hets deals with multiple
different logics seamlessly by treating translation from one logic to
another (comorphisms) as a first-class citizen. Among the logics
supported by Hets are CoCASL (a coalgebraic extension of CASL),
ModalCASL (an extension of CASL with multi-modalities and term
modalities) and Isabelle/HOL\,\cite{DBLP:books/sp/NipkowPW02} (an
interactive theorem prover for higher-order logic).

Although ITP, CITP, and InvA can automate theorem proving to some
extent, human interaction with theorem proving is mandatory. Thus, the
tools are interactive theorem provers or proof assistants. Among the
other proof assistants are ACL2\,\cite{KaufmannMM2000ACL2},
Coq\,\cite{DBLP:series/txtcs/BertotC04},
Isabelle/HOL\,\cite{DBLP:books/sp/NipkowPW02},
HOL4\,\cite{DBLP:conf/tphol/SlindN08},
HOL Light\,\cite{DBLP:conf/fmcad/Harrison96},
LEGO~\cite{Luo92},
NuPRL~\cite{nuprl},
Agda\,\cite{DBLP:conf/tphol/BoveDN09},
Lean~\cite{DBLP:conf/itp/NawrockiAE23,DBLP:conf/itp/AyersJG21},
and
PVS\,\cite{DBLP:conf/cade/OwreRS92}. The proof assistants have notable
applications in industry as well as academia. One distinguished case
study conducted with ACL2 is a mechanically checked proof of the
floating point division microcode program used on the
AMD5\({}_{\mbox{K}}\)86
microprocessor\,\cite{DBLP:journals/tc/MooreLK98}. The proof was
constructed in three steps: (1) the divide microcode was translated
into a formal intermediate language, (2) a manually created proof was
transliterated into a series of formal assertions in ACL2, and (3)
ACL2 certified the assertion that the quotient will always be
correctly rounded to the target precision. 
There are several recent verification case studies using ACL2.
Hardin~\cite{DBLP:journals/corr/abs-2311-08862} used it to formally verify his implementation of the Dancing Links optimization, an algorithm proposed by Knuth in his book~{\it The Art of Computer Programming}~\cite[Volume 4B]{knuth-taocp} to provide efficient element removal and restoration for a doubly linked list data structure.
The data structure was implemented in the Restricted Algorithmic Rust~(RAR), a subset of the Rust programming language crafted in his prior work~\cite{DBLP:journals/corr/abs-2205-11709}.
He then used his RAR toolchain to {\it transpile} (i.e., perform a source-to-source translation) the RAR source into the Restricted Algorithmic C~(RAC)~\cite{DBLP:books/sp/Russinoff22}.
The resulting RAC code was converted to ACL2 by leveraging the RAC-to-ACL2 translator~\cite{DBLP:books/sp/Russinoff22} and the formal verification was finally conducted.
Gamboa et al.~\cite{DBLP:journals/corr/abs-2311-08857} presented a tool that can suggest to ACL2 users additional hypotheses based on counterexamples generated when the theorem under verification does not hold as expected with the aim of making the theorem become true.
Kumar et al.~\cite{DBLP:journals/corr/abs-2311-08859} formalized the GossipSub peer-to-peer network protocol popularly used in decentralized blockchain platform, such as Ethereum 2.0, and verified its security properties.

Coq was used in the
CompCert project\,\cite{DBLP:journals/cacm/Leroy09} in which it has
been formally verified that an optimizing C compiler generates
executable code from source programs such that the former behaves
exactly as prescribed by the semantics of the latter. Coq was also
used to formally prove the famous four-color
theorem\,\cite{Gonthier2008NAMS}.  
Making good use of Coq, the Iris separation logic framework~\cite{DBLP:journals/jfp/JungKJBBD18} has emerged as an effective way to reason about concurrent programs in the context of recent advances in compilers and computer processors.
Jung et al.~\cite{DBLP:journals/pacmpl/JungLCKPK23} used Iris to verify some concurrent data structures where {\it safety memory reclamation} was taken into account, i.e., an unused memory block by a thread can be safely freed with a guarantee that no other thread accesses such a freed memory block.
M{\'{e}}vel et al.~\cite{DBLP:journals/pacmpl/MevelJ21} verified the safety of
a concurrent bounded queue with respect to a weak memory model, i.e., the compiler and/or processor can reorder the load and store operations of a single thread to improve the overall performance as long as the reordering does not affect the behavior of that thread.
Iris was also applied in an industrial context where some data structures used in Meta were verified~\cite{DBLP:conf/cpp/CarbonneauxZKON22,DBLP:conf/cpp/VindumFB22}.
Some other recent studies used Coq to reason about probabilistic programs~\cite{DBLP:conf/cpp/AffeldtCS23,DBLP:journals/pacmpl/ZhangA22}, 
smart contracts~\cite{DBLP:conf/cpp/NielsenAS23,DBLP:conf/cpp/AnnenkovMNS21,DBLP:conf/cpp/AnnenkovNS20}, and
cryptographic protocols in the game-based model~\cite{DBLP:journals/toplas/HaselwarterRMWASHMS23}.

HOL4 and HOL Light are two members of the HOL theorem prover family. 
HOL4 is designed to be a high-performance, feature-rich environment, and capable of handling sophisticated proofs and formalization problems.
In contrast, HOL Light, which was mainly developed by John Harrison, focuses on simplicity and ``minimalism.''
While HOL4 offers an extensive feature set and tools for interactive and automated theorem proving, HOL Light aims to provide a clean and straightforward system for higher-order logic proofs, focusing on ease of use and conceptual clarity.
The HOL family provers and Isabelle/HOL have been used to verify 
the IEEE 754 floating-point standard\,\cite{DBLP:journals/fmsd/Harrison00} and
the Kepler Conjecture\,\cite{hol_Kepler_conjecture}.

Some proof assistants allow the extraction of executable programs from their formal specifications.
Agda supports generating Haskell programs, while Coq supports generating OCaml and Scheme as well as Haskell programs.
However, the generated programs are functional, not imperative or concurrent.
A technique for translating CafeOBJ specifications describing OTS to Java concurrent programs has been proposed\,\cite{DBLP:conf/apsec/HaO17}, but human users need to write the program manually. The technique only inserts some annotations into the specification to guide humans, who are assumed to not understand the CafeOBJ specifications, to write corresponding Java programs. 
For example, an annotation can let human user know that the associated code block should be run on a separate thread.
Another study\,\cite{DBLP:conf/seke/SenachakSOF05} has proposed a method and tool for generating Java sequential programs from OTS/CafeOBJ specifications.
In summary, for concurrent and distributed systems, there is still a gap between formal specifications used for formal verification and the corresponding implementations.
Verifying programs directly in their implementation languages like Java without translating or specifying them in another formal specification language is a possible way to mitigate this gap.
KeY~\cite{DBLP:series/lncs/10001} makes it possible to do so, though only Java sequential programs are supported.
The Java Modeling Language (JML) is used to specify desired properties with preconditions, postconditions, and invariants of the program under verification.
Note that multithreading programs are not supported.
KeYmaera X~\cite{DBLP:conf/cade/FultonMQVP15} is the hybrid version of KeY, designed for verifying hybrid systems.
Although some case studies have demonstrated that it is possible to verify hybrid systems with the proof score approach\,\cite{DBLP:conf/ifip10-3/OgataYSF04,DBLP:journals/corr/abs-2010-15280}, the approach is not specifically designed for such systems.

Among formal methods along the line of one main stream are
Z\,\cite{DBLP:books/daglib/0072139} and
Event-B\,\cite{DBLP:books/daglib/0024570}
in which stepwise
refinement plays the central role. These formal methods are equipped
with proof assistants or environments in which formal proofs related
stepwise refinement are supported.  For example,
Z/EVES\,\cite{DBLP:conf/zum/Saaltink97} is a proof assistant for Z,
Rodin\,\cite{DBLP:journals/sttt/AbrialBHHMV10} is an open toolset for
modeling and reasoning in Event-B in which a proof assistant is
available. Event-B can be also verified with the ProB model 
checker~\cite{event-b-prob}, which is integrated into Rodin.
Stepwise refinement
or simulation-based verification can be conducted for OTSs in
CafeOBJ\,\cite{DBLP:journals/entcs/OgataF08,mcs_access}, but such formal
verification is not supported by CiMPA and CiMPG. It is one possible
direction to extend CiMPA and CiMPG so as to support stepwise
refinement or simulation-based verification for OTSs.

SAT/SMT solvers and SAT/SMT-based formal verification techniques/tools
have been intensively studied because they make it possible to
automate formal verification experiments. Among SAT/SMT solvers are
Z3\,\cite{DBLP:conf/tacas/MouraB08} and
Yices\,\cite{DBLP:conf/cav/Dutertre14}. The standard input format to
SAT/SMT solvers is conjunctive normal form (CNF). Because CNF is not
very user friendly, many other tools/environments have been developed
in which SAT/SMT solvers are used internally. Among such
tools/environments are Dafny\,\cite{DBLP:conf/lpar/Leino10}, Symbolic
Analysis Laboratory (SAL)\,\cite{DBLP:conf/cav/MouraORRSST04} and
Why?\,\cite{DBLP:conf/esop/FilliatreP13}. These tools/environments
have high-level specification languages in which systems are specified
and user interfaces for formal verification used by SAT/SMT
solvers. Z3 is used by Dafny and Why?, while Yices is used by SAL and
Why?. Why? also uses PVS, Isabelle/HOL, Coq, and many other automated
theorem provers, such as Vampire\,\cite{DBLP:conf/cav/KovacsV13} and E
prover\,\cite{DBLP:conf/cade/0001CV19}.

SAT/SMT solvers have been integrated into or combined with some
existing formal methods tools/environments. 
ALC2 was
extended with SMT solvers\,\cite{DBLP:journals/corr/PengG15}.
SMTCoq\,\cite{DBLP:conf/cav/EkiciMTKKRB17} is an open-source plugin
for Coq that dispatches proof goals to external SAT/SMT. When such
solvers successfully prove a goal, they are supposed to return a proof
witness, or certificate, which is then used by SMTCoq to automatically
reconstruct a proof of the goal within Coq. 
Barbosa et al.~\cite{DBLP:conf/lpar/BarbosaK0VTB23} recently extended SMTCoq with a newly interactive tactic called \texttt{abduce}.
When the SMT solver fails to prove a goal $G$ valid under the given hypotheses $H$, instead of returning a counterexample, this new tactic exploits the  abductive capabilities of the CVC5 SMT solver~\cite{DBLP:conf/tacas/BarbosaBBKLMMMN22} to produce some additional assumptions (formulas) that are consistent with $H$ and make $G$ provable.
Rocha et\ al.\ proposed
rewriting modulo SMT\,\cite{DBLP:journals/jlp/RochaMM17} that is a
combination of SMT solving, rewriting modulo theories, such as
associativity and commutativity, and model checking. Rewriting modulo
SMT is suited to model and analyze reachability properties of
infinite-state open systems, such as cyber-physical systems.

Some recent work investigated the development of graphical user interfaces for the existing proof assistants, such as those for Coq~\cite{Donato2023IntegratingGP,DBLP:conf/cpp/DonatoSW22} and those for Lean~\cite{DBLP:conf/itp/NawrockiAE23,DBLP:conf/itp/AyersJG21}.
Korkut~\cite{DBLP:journals/corr/abs-2303-05865} developed a web-based graphical proof assistant - Proof Tree Builder, from which users can construct proofs for simple imperative programs based on Hoare logic~\cite{DBLP:journals/cacm/Hoare69}.
Instead of writing textual proof commands, 
users of those tools can perform gestural actions like click and drag-and-drop on the graphical interface with terms representing, for example, the current goal, to construct the proof.
The authors of those studies argued that this can provide an intuitive and quick way of proof construction.

The proof score approach to formal verification is unique in that (1)
proofs (or proof scores) are written in the same language, CafeOBJ in
this paper, as the one for systems and property specification and (2)
proof scores can be written as programs and then are flexible as
programs. However, proof scores are subject to human errors because
proof scores basically need to be written by human users as
programs. To overcome this weakness, CiMPA and some other proof
assistants for CafeOBJ have been built, although these proof
assistants dilute the merits of proof scores. To address this issue,
CiMPG has been developed. CiMPG overcomes the weakness of proof scores
and also keeps the merits of proof scores. One possible evolution of
CiMPA and/or CiMPG is to integrate SAT/SMT solvers into them. This
would be feasible because rewriting modulo SMT is available in
Maude. It is, however, necessary to consider how to use SAT/SMT
solvers in CiMPA and/or CiMPG so that the merits of proof scores can
be enjoyed.
Finally, the generate \& check method~\cite{FUTATSUGI2022102893}
extend the proof score methodology by combining reductions and searches,
which gives hints for future developments of the proof score technique.

\textbf{Discussion: what features have contributed to the success of other tools?}
It is worth discussing why some tools have been more widely adopted than proof
scores. We will focus on Event-B, Isabelle/HOL, Coq, and Lean.

As we sketched above, Event-B focuses on refinements to gradually add
details and functionality to systems. The specifier starts with a very general
system and introduces, step by step, improvements and components until
the behavior reaches the desired level. Proof obligations are generated for
the first specifications and for every enrichment step; automation is supported
by tools like Rodin and ProB, introduced above. This process eases verification,
as each step should be easier to verify than the complete system.
It is also relevant the effort made by the community around the ABZ conference,
which proposes case studies for specification and analysis in Event-B.
The research in papers like~\cite{event-b-automotive,event-b-lung} is
illustrative of the discussion above: the case studies, the adaptive exterior light 
system for cars and the model of a mechanical lung ventilator, 
were proposed in the ABZ conference (in the 2020 edition and 2024 edition, 
respectively). Both were modeled and progressively refined, had their proof
obligations discharged by using Rodin and the model was validated by using
the ProB model checker.

Coq and Lean are proof assistants based on the proposition-as-types paradigm
of the calculus of inductive constructions. Historically, Coq has been
used for verifying a wide range of systems
thanks to its expressive logic, its high level of automation, and mature ecosystem,
which includes plug-ins for Visual Studio, Emacs, and Vim. Recent studies include
specifications and proofs for systems as different as the Ethereum Virtual 
Machine~\cite{AlbertGKM23} and the semantics of quantum languages~\cite{ShiCD24}.
Lean, developed by Microsoft, inherits the philosophy of Coq. Although this theorem
prover is not as mature as Coq yet, it has a very active community. It is also integrated
with Visual Code and most of the current efforts (e.g.~\cite{CarneiroBU23,abs-2403-14064}) 
are focused on the mathlib library, an in-development unified library of mathematics.

Isabelle/HOL, possibly the most used higher-order logic theorem prover, has a number
of advantages: (i) versatile user interface, including integration with
Visual Studio and even an online coding platform, Isabelle/Cloud~\cite{XuZ23};
(ii) powerful automation by means of Sledgehammer~\cite{sledgehammer}, which tries to
discharge automatically the current goal by invoking automatic theorem provers; 
(iii) human-readable proof scripts, written using the Isar (Intelligible semi-automated
reasoning) language~\cite{isar}; and
(iv) a strong community and active development, being used nowadays for 
analyzing, for example, memory access violations~\cite{AhmadiDG24} and
cyber-physical systems~\cite{MuniveFGSLH24}.

\textbf{Why using proof scores?}
The discussion above showed some weak points of proof scores, mainly its lack
of advanced user interfaces, which has been also discussed in Section~\ref{sec:open}.
We also find inspiration for some issues:
\begin{itemize}
\item
New protocols can be
obtained from challenges like the ones proposed by others, including the ABZ
conference.

\item
Software engineers do not proceed as researchers. Extending proof scores
with features for refinement (already explored in CafeOBJ, as discussed above)
and producing easier, human-readable proof (like Isar) would make proof
scores more attractive and resolve (even if only partially) the need for experts

\item
Each proof starts from scratch, without libraries for easing proofs in certain
contexts. Developing libraries and making them public would attract new
users and ease automatization, hence being a piece of future work.
\end{itemize}

Moreover, we can argue that proof scores equal, or even surpass, several of the
aspects described above:
\begin{itemize}
\item
A very expressive syntax, especially including the support for equational axioms,
such as commutativity and associativity.

\item
The operational semantics, as described in Section~\ref{sec:theory}, are simple
but very powerful.

\item
An efficient C++ implementation (for those Maude-based implementations, see
Section~\ref{subsec:comparison}). 

\item
These three aspects together (expressive
syntax, powerful and simple semantics, and efficient implementation) make proof
scores very convenient. On the one hand, because we use the same language
for specifying and proving, the more powerful is the language, the simpler the
proof will be.

On the other hand, only a few languages can handle efficiently equational
axioms. There exist translations from Maude into Isabelle/HOL~\cite{hets-amast10}
and from Maude into Lean~\cite{RubioR25}. In both cases, the specifications and
the proofs are more verbose in the target language than in Maude and the axioms
need to be used carefully by hand to avoid infinite loops.
This expressivity can be further improved by using symbolic analysis.

\item
The automation level reached by CiMPG+F (see Section~\ref{subsubsec:cimpg})
is comparable to Sledgehammer, as shown in~\cite{DBLP:journals/jss/RiescoO22}.
Although future development should improve this tool, its current capabilities
are state-of-the-art.

\item
The Maude-based implementations (see Section~\ref{sec:tools}) support
model checking of LTL formulas. The automatic nature of the model checker
makes it very adequate for software engineers, allowing them to prove
properties (e.g.\ safety properties, which are recurrent in industry) more easily.
Moreover, CiMPG (see Section~\ref{subsubsec:cimpg})
provides a \texttt{proven} command~\cite{DBLP:journals/jss/RiescoO22}
for asserting properties; connecting both tools is an interesting topic of future work.

\item
Although many proofs have been developed several years ago, recent advances
in post-quantum protocols have been done recently, as discussed in 
Section~\ref{sec:open}. As discussed above, more
modern challenges should be achieved in order to attract new researchers, but
these protocols prove that the proof score methodology is still relevant nowadays.
\end{itemize}
 
\section{Conclusions}\label{sec:conc}

In this paper we have presented proof scores, a verification methodology
that allows users to prove properties in systems in a flexible way.
We have first defined the theoretical framework and the different tools
supporting the approach, including those implementing novel features that
ease the verification process. Then, we have analyzed the strong points
of the approach, illustrated in many case studies that have been
successfully analyzed, but also the weak points and open issues that 
should be addressed in order to make proof scores a widely used technique.

It is worth summarizing here how these open issues should be addressed.
First, most modern systems provide an IDE and hence it is worth providing
graphical support for proof scores. In this sense, it should help the
user for both developing the specification and the proof. Because some
of the latest CafeOBJ interpreters support interactive theorem proving
and it is able to relate it to proof scores it would be worth integrating
a graphical representation of the proof, so the user is aware of the
remaining subgoals. 
Moreover, it would be nice to integrate graphical tools that help the
user to find the most appropriate case splittings, possibly showing
different runs of a (non-deterministic) protocol and presenting those
terms that remain invariant and those that change.

Then, it is worth exploring how new Maude features can be used in theorem
proving. First, it would be interesting to analyze how symbolic analysis,
using unification and narrowing, can be applied in proof scores.
Moreover, the latest Maude release supports meta-interpreters, which use
Maude meta-level and work as a standard Maude terminal, supporting execution
of terms in different processes. Using these features, we could even use
potentially non-terminating computations in our proofs: they would be
executed in a different process while the main proof continues by analyzing
other subgoals; if the analysis does not finish after some time (chosen
by the user), then the computation can be stopped and other analysis can
be tried.

Regarding the properties that can be proven using proof scores,
it is interesting to explore how to tackle liveness properties. In this
case we should consider using a co-algebraic approach (or hidden algebra or behavioral
specifications), because we need to have infinite sequences of states.

All these features should be applied to novel protocols, like quantum
and blockchain protocols. Using proof scores for proving properties for these
systems will possibly lead to new techniques, illustrate its power, and
set the foundations for the future of proof scores.

{
\bibliographystyle{ACM-Reference-Format-Journals}

}

\received{Month Year}{Month Year}{Month Year}

\end{document}